\documentclass[amsmath,amssymb,pre,twocolumn,longbibliography]{revtex4-1}
\usepackage{bookmark}
\usepackage{graphicx}% Include figure files
\usepackage{dcolumn}% Align table columns on decimal point
\usepackage{subfigure}
\usepackage{bm}% bold math
\usepackage{color}
\usepackage{braket}

\newtheorem{thm}{Theorem}

\begin{document}

\title{Branches of triangulated origami near the unfolded state}

\author{Bryan Gin--ge Chen}
\author{Christian D. Santangelo}
\affiliation{Department of Physics, University of Massachusetts Amherst, Amherst, MA  01003}

\begin{abstract}
Origami structures are characterized by a network of folds and vertices joining unbendable plates. For applications to mechanical design and self-folding structures, it is essential to understand the interplay between the set of folds in the unfolded origami and the possible 3D folded configurations. When deforming a structure that has been folded, one can often linearize the geometric constraints, but the degeneracy of the unfolded state makes a linear approach impossible there. We derive a theory for the second-order infinitesimal rigidity of an initially unfolded triangulated origami structure and use it to study the set of nearly-unfolded configurations of origami with four boundary vertices. We find that locally, this set consists of a number of distinct ``branches'' which intersect at the unfolded state, and that the number of these branches is exponential in the number of vertices.  We find numerical and analytical evidence that suggests that the branches are characterized by choosing each internal vertex to either ``pop up'' or ``pop down''.  The large number of pathways along which one can fold an initially-unfolded origami structure strongly indicate that a generic structure is likely to become trapped in a ``misfolded'' state. Thus, new techniques for creating self-folding origami are likely necessary; controlling the popping state of the vertices may be one possibility. 
\end{abstract}
\date{\today}

\maketitle

\section{Introduction}
The development of responsive materials has paved the way to the fabrication of self-folding structures, based on origami, in which flat sheets of a material can be folded along a discrete network of creases into a targeted three-dimensional configuration \cite{silverberg2015origami,metasheets,na2015programming,sussman2015,ryu2012,tolley2014,liu2014,tachi2016self}. The creases and vertices formed at their junctions together form a kind of geometric ``program'' which determines the shape from the strong constraints on how a flat sheet can fold into space. This attractive design paradigm suggests the use of origami as the foundation for mechanical metamaterials \cite{Wei2013,Silverberg2014,tachigeometric,Evans2015,Lv2014,schenk2013} and deployable structures \cite{miura,pellegrinobook}. Yet, flexibility is both a blessing and a curse: a single origami crease pattern can admit many different folding pathways \cite{hull2002combinatorics,ballinger2015minimum,abel2015rigid} and, indeed, manipulating a nearly-unfolded origami structure with one's hands (e.g.~the ``map-folding problem'') illustrates the competition between pathways that can impede folding to a specific desired configuration \cite{Silverberg2014,metasheets,tachi2016self,brunck2016}. Furthermore, experiments on self-folding gel origami do not always fold into the expected, programmed shape \cite{silverberg2015origami}.

Let us now fix terminology so that we can discuss these issues more precisely. An {\em origami structure} refers to a system of rigid flat plates joined pairwise by ideal hinges, or {\em creases}. Origami structures considered here will always arise from a plane polygon decorated with a network formed by the creases and their junctions at vertices, which we call the {\em crease pattern}.  Origami structures can take on a variety of {\em configurations} in 3D space, which are uniquely specified by the positions of their vertices. 

\begin{figure}[b]
\includegraphics[width=3.5in]{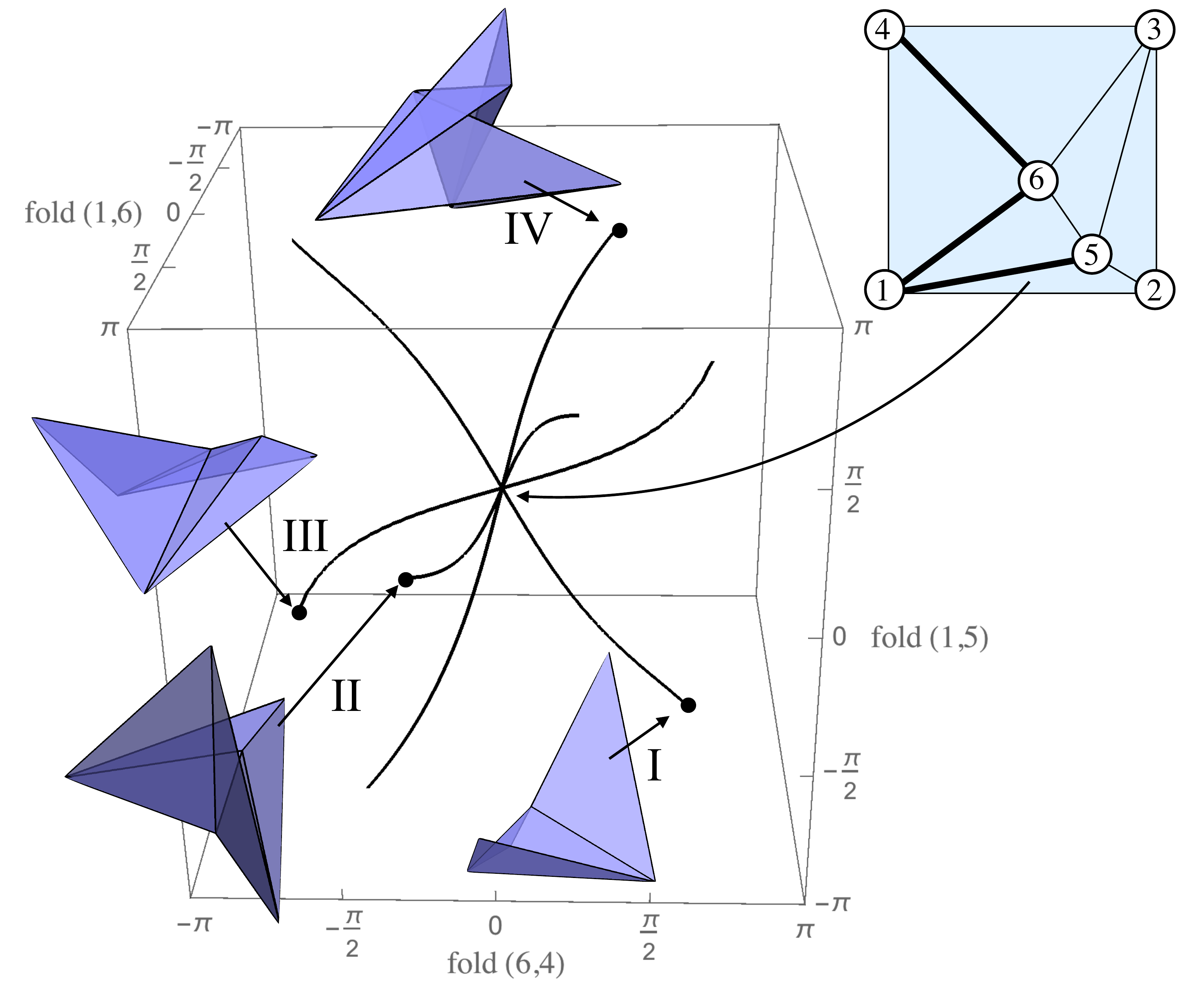}
\caption{\label{fig:schematic} A neighborhood of the unfolded state in the configuration space of a two-vertex origami structure (inset) projected onto the fold angles of 3 folds (thick lines). This was computed by solving numerically the length constraint equations (Eq.\ \ref{eq:length}). Locally, there are 4 branches, labeled I-IV, each a one-dimensional configuration space, all intersecting at a single point: the flat, unfolded configuration. Supplementary Movies 1-4 show animations of these branches. We did not attempt to compute the global structure of the configuration space, e.g.\ how the branches join.}
\end{figure}

To better understand the phenomena of multiple folding pathways and misfolding, it is useful to distinguish two notions of floppiness in an origami structure: (1) the number of degrees of freedom, $D$, which is the dimensionality of the space of motions and scales with the number of boundary sides of a generic origami crease pattern \cite{tachigeometric,chen2016topological,santangelo2017extreme}, and (2) the number of distinct {\em branches} $B$, or folding pathways. As we discuss later, the flat unfolded configuration of an origami structure is a singularity in the space of origami configurations where $B$ branches of dimension $D$ intersect. Consider the triangulated origami structure in Fig.\ \ref{fig:schematic} as an example; the figure shows some of the allowed configurations as a function of three of the fold angles. In this example, $D=1$ so the configuration space is locally curve-like almost everywhere; yet, $B=4$, which can be seen as four distinct one-dimensional branches intersecting at the unfolded configuration in the center. 

This paper addresses the following questions: how many distinct branches does a generic, triangulated origami crease pattern have, and how can those branches be distinguished? We focus on triangulated crease patterns because triangulated structures are marginally rigid while being maximally flexible \cite{chen2016topological}, and because triangulated origami can encode the kinematics of origami with bendable faces more generally \cite{Silverberg2014}. 

To answer these questions, we first show that the small deformations around any initially unfolded configuration of a triangulated origami structure can be described by the simultaneous solutions of a system of $V_i$ quadratic equations (Section \ref{sec:methods}) in the $V_i+V_e$ vertical displacements at the vertices, where $V_i, V_e$ are the number of internal and external vertices, respectively. We will give two interpretations for each of these equations, one coming from statics, showing that we have one equation for each self stress in the system, and one coming from kinematics, showing that each equation enforces the vanishing of Gaussian curvature at an internal vertex. We use this formalism to review the geometry of nearly-unfolded $n$-fold single-vertex origami structures and give a new proof of the fact that their configuration spaces look like $(n-3)$-dimensional double cones \cite{kapovichmillson}, where the two nappes are distinguished by whether the vertex ``pops'' up or down \cite{abel2015rigid,tachi2016self}. 

Moving on to the case of triangulated origami with multiple vertices, we restrict our attention in this paper to triangulated origami with four boundary vertices, where the number of degrees of freedom $D=1$. We provide in Section \ref{sec:numerical} numerical evidence from a model of random origami squares that the number of branches, $B$, is \textit{generically} $2^{V_i}$, and they are with high probability all distinct. We find a small number of exceptions (appearing with frequency $\sim 1/1000$) that can all be identified from the crease pattern. The branches are not necessarily distinguished by the mountain and valley assignments of the folds, that is, which folds have dihedral angle larger or smaller than $\pi$ respectively. However, we find that pairs of branches appear to be in one-to-one correspondence with pairs of {\em vertex sign patterns}, which are assignments of $\pm1$ to each internal vertex specifying their popping state. 

In Section \ref{sec:butterfly} we show that a special class of triangulations (roughly, those formed from a sequence of adding degree-3 vertices to the boundary) do satisfy $B=2^{V_i}$. We conclude with a discussion in Section \ref{sec:discussion} with a discussion on the implications of our results. In particular, we note that the exponential number of branches of a generic origami crease pattern interferes with typical designs for self-folding origami, thus requiring a deeper understanding of how to engineer the origami configuration space topology. Specifically, our results suggest that methods for controlling the popping state of vertices should be investigated.

\section{Analytical Methods}\label{sec:methods}
\subsection{Model and second-order deformations}
\label{subsec:model}

\begin{figure}[b]
\includegraphics[width=3.5in]{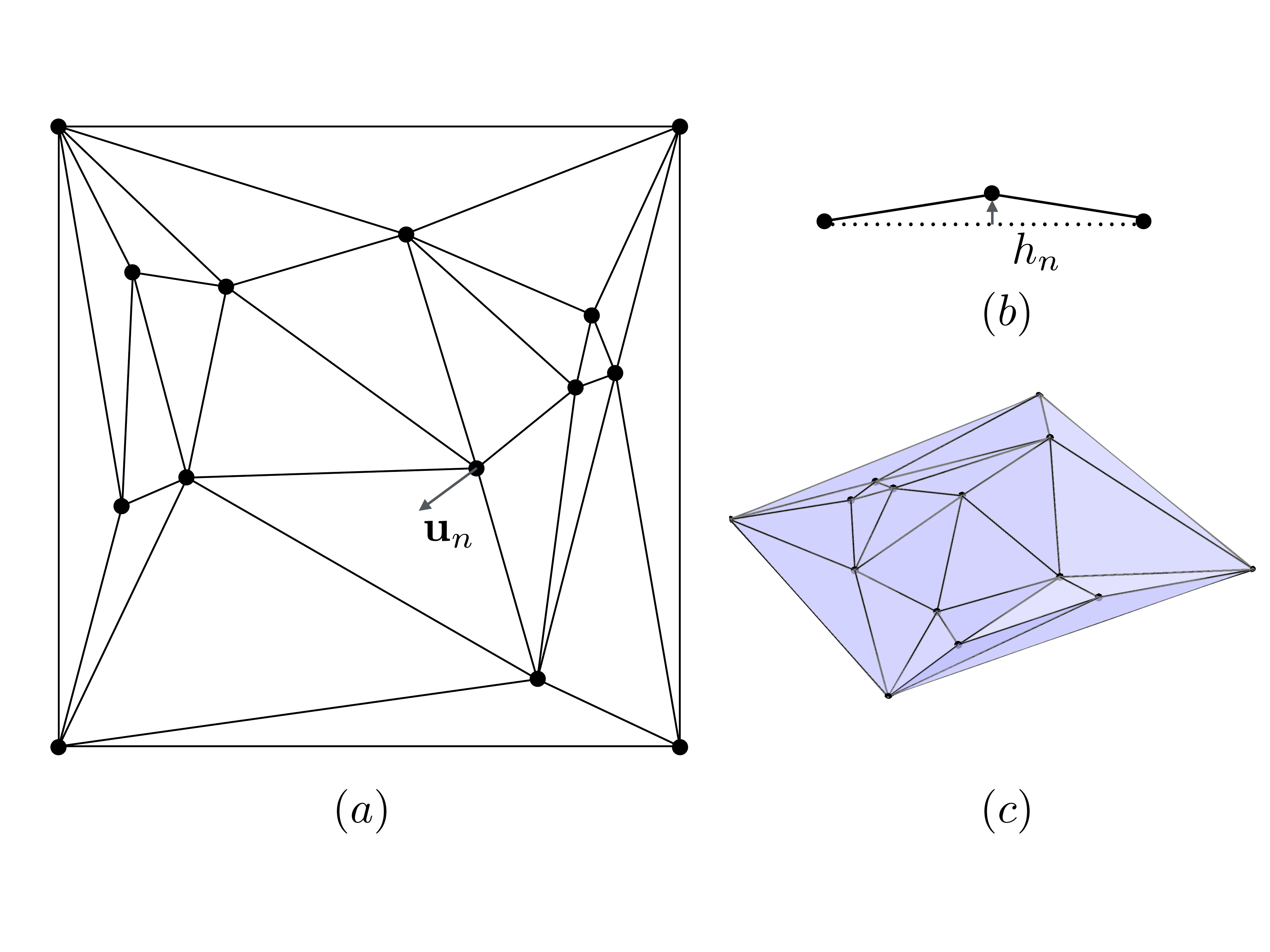}
\caption{\label{fig:notation} (a) An origami in the unfolded state; the vector ${\bf u}_n$ represents the in-plane displacement of vertex $n$. (b) A cross-sectional side view near vertex $n$; the scalar $h_n$ gives the vertical displacement of vertex $n$. (c) A small deformation of this origami away from the unfolded state.}
\end{figure}

Our kinematic model for origami consists of a triangulated network of springs joining vertices in two dimensions which can, upon deformation, come out of the plane (Fig.\ \ref{fig:notation}). We will only consider networks that are planar triangulations of disks (polygons). The edges that are in the interior of the disk will be called {\em folds}, since we think of the network as a representation of an origami crease pattern, and since they separate pairs of triangles whose relative orientations differ by some dihedral angle at that edge. We refer to configurations of the origami structure as {\em folded} if not all of the dihedral angles between adjacent faces are equal to $\pi$, and {\em unfolded} otherwise. Because the network is made from triangles, the angles at vertices of faces between adjacent edges will be preserved as long as lengths are preserved. Furthermore, the triangular faces cannot bend, making this a good model for rigid origami. We label the vertices of the origami structure by an integer, and label edges by a pair $(n,m)$ when the fold joins vertex $n$ to vertex $m$. Using this notation, the kinematic constraints are given by the equations,
\begin{equation}\label{eq:length}
|\mathbf{X}_n - \mathbf{X}_m|^2 - L_{(n,m)}^2 = 0,
\end{equation}
for each pair of vertices, $(n, m)$, joined by a spring of equilibrium length $L_{(n,m)}$.  These equations define the {\em configuration space} $\mathcal{C}$ of the origami. Note that while self-intersections are allowed, we will work in neighborhoods of $\mathcal{C}$ consisting of configurations which do not self-intersect.

Fig.\ \ref{fig:schematic} shows numerical solutions to these length equations for a simple crease pattern. We have chosen to draw $\mathcal{C}$ in this figure using fold angles as coordinates here since those variables are more intrinsic (for instance, they do not see the overall position and orientation of the structure). Near the unfolded state, which we take as the origin, one can linearize the fold angles as functions of the displacements. Therefore the shape of $\mathcal{C}$ near the origin looks the same (up to this linear map) in either set of coordinates. This representation is also useful when thinking about the self-folding paradigm, which we will return to in the concluding section. 

The two notions of floppiness described in the introduction have natural interpretations in terms of the geometry of $\mathcal{C}$. The number of degrees of freedom $D$ is the dimension of the configuration space. Note that the configuration space may have singularities (and in this work, this is the case of particular interest), so the notion of ``dimension'' becomes subtle (\cite{harris}, Lecture 11). For our purposes, it will suffice to say that the dimension of the configuration space is the dimension at any nonsingular point.

The number of branches $B$ is a property of a singular point of $\mathcal{C}$. For instance, at a singular point consisting of the intersection of multiple distinct curves or surfaces, each one of those would consist of a branch. A general definition of a branch (as an irreducible component of the analytic germ at the singularity) would take us a bit too far afield into singularity theory \cite{greuel}; we give a more concrete definition for our case later in Section \ref{subsec:count}.

We are interested in deformations of unfolded origami, where the vertices all lie in a single plane and the faces do not overlap. Without loss of generality, we will assume the initial unfolded configuration lies in the $xy-$plane. We write the position of vertex $n$ as $\mathbf{X}_n = \mathbf{U}_n + \mathbf{u}_n + h_n \mathbf{\hat{\mathbf{z}}}$, where $\mathbf{U}_n$ are the equilibrium positions of the vertex in the $xy-$plane, $\mathbf{u}_n$ is a vertex displacement in the $xy-$plane and $h_n$ is a vertical displacement out of the plane. Expanding Eq.\ (\ref{eq:length}) to lowest order in the displacements yields
\begin{equation}\label{eq:expanded}
2 \frac{\mathbf{U}_n - \mathbf{U}_m }{\left|\mathbf{U}_n - \mathbf{U}_m \right|} \cdot \left(\mathbf{u}_n - \mathbf{u}_m\right) + \frac{\left(h_n-h_m \right)^2}{\left|\mathbf{U}_n - \mathbf{U}_m \right|} \approx 0.
\end{equation}
Because the vertical displacement decouples from the in-plane displacement and the linear terms in $h_n$ vanish, any displacement with $\mathbf{u}_n=0$ for all $n$ preserves lengths to first order (i.e.~any displacement consisting only of height changes is a first-order flex or motion). But by stopping here we have not captured enough information to see the branches, as the lowest-order information lies in the quadratic terms of Eq.\ (\ref{eq:expanded}). Since the term quadratic in height leads to a change in bond lengths of the same order as the linear term in the in-plane displacements, we can safely neglect terms of order $\mathcal{O}(\mathbf{u}_n^2)$.

The first term of Eq.\ (\ref{eq:expanded}) governs the infinitesimal displacements of the in-plane degrees of freedom. We can rewrite this expression by concatenating the in-plane displacements into a vector $(\mathbf{u}_1, \mathbf{u}_2,\cdots)$, and define an in-plane compatibility matrix such that row $(n,m)$ of $\mathbf{C}$ is defined by the equation
\begin{equation}\label{eq:isometry}
\left[\mathbf{C} \left(\begin{array}{c}
\mathbf{u}_1\\
\vdots
\end{array}\right)\right]_{(n,m)} = \frac{\mathbf{U}_n - \mathbf{U}_m }{\left|\mathbf{U}_n - \mathbf{U}_m \right|} \cdot \left(\mathbf{u}_n - \mathbf{u}_m\right).
\end{equation}
The matrix $\mathbf{C}$ has pairs of columns indexed by vertices $n$ and rows indexed by the unique folds $(n,m)$. Formally, it maps vectors of in-plane deformations to vectors of in-plane spring displacements and governs the linear deformations of the unfolded configuration of the origami structure that keep it in the $xy$-plane. Using $\mathbf{C}$, Eq.\ (\ref{eq:expanded}) becomes
\begin{equation}\label{eq:expanded2}
2[\mathbf{C}\cdot\mathbf{u}]_{(n,m)}+ \frac{\left(h_n-h_m \right)^2}{\left|\mathbf{U}_n - \mathbf{U}_m \right|} \approx 0.
\end{equation}

Since the in-plane deformations and out-of-plane deformations are decoupled in Eq.\ (\ref{eq:expanded2}), the in-plane motions are governed (at this order) by the kernel of $\mathbf{C}$. And because 2D triangulated networks are generically rigid in the plane, by a counting argument and Laman's theorem \cite{graverservatiusservatius}, $\textrm{ker} ~\mathbf{C}$ is generically generated by translations and rotations only. As these are not of much interest to us here, from now on, we will only consider how Eq.\ (\ref{eq:expanded2}) constrains out-of-plane deformations.

To extract constraints on the vertical displacements $h_n$ from Eq.~(\ref{eq:expanded2}), we make use of the {\em self stresses} of the network. These are {\em row} redundancies in $\mathbf{C}$, defined by $\bm{\sigma}^T \cdot \mathbf{C} = \mathbf{0}^T$, where $\bm{\sigma}$ is a stress vector with one component per fold and $^T$ is the transpose \footnote{Note that unlike the dimension of the space of linear deformations, the space of self stresses does not depend on whether the unfolded configuration is considered to be embedded in an ambient 2D or 3D space.}. Taking the dot product of $\bm{\sigma}$ with both sides of Eq.\ (\ref{eq:expanded}), we obtain
\begin{equation}\label{eq:compat}
\sum_{(n,m)} \frac{\sigma_{(n,m)}}{|\mathbf{U}_n-\mathbf{U}_m|} \left(h_n-h_m\right)^2 = 0,
\end{equation}
where $\sigma_{(n,m)}$ is the component of $\bm{\sigma}$ along the fold $(n,m)$ and the sum is taken over all folds in the network. Thus, each self stress of the unfolded configuration of the origami gives an equation (\ref{eq:compat}) that constrains the vertical displacements $h_n$ to second order. This is a special case of the more general formalism of Connelly and Whiteley \cite{connellywhiteley} who also show that all second-order constraints are generated by the self stresses (that is, these conditions which are necessary for deformations to preserve lengths to second-order are also sufficient).

In most of the rest of this paper, we will be considering solutions to the system of equations coming from applying Eq.\ (\ref{eq:compat}) to an independent basis of self stresses of triangulated crease patterns. This system defines a certain subspace in the space of all vertical displacements, whose dimension we will discuss in Section \ref{subsec:count}. By considering only vertical displacements as our variables we are implicitly removing the in-plane degrees of freedom, and in particular, the in-plane translations and rotation about the $z$-axis will not come into play. 

The solutions $h_n$ to Eq.~(\ref{eq:compat}) only give an approximation to the configuration space at the unfolded state, and may not be a wholly accurate picture, even qualitatively (\cite{harris}, Lecture 20). The issue is that while the non-existence of folded solutions at second-order proves that there can be no folded configurations (second-order rigidity implies rigidity) \cite{connellywhiteley,connellycabled}, a second-order solution may not be a solution at all orders. In all cases that we checked, e.g.~in making Fig.~\ref{fig:schematic} and the Supplementary Movies, solutions to Eq.~(\ref{eq:compat}) did seem to correspond to true solutions of Eq.~(\ref{eq:length}) (see Section \ref{sec:numerical}). And for single-vertex origami, the second-order deformations from the unfolded state can be shown to extend to actual rigid motions \cite{kapovichmillson,reciprocalarea}, see also Appendix \ref{app:quadratic}). We also discuss in Section \ref{sec:butterfly} some special multiple-vertex crease patterns where second-order motions also extend to true motions. However, we know of no such general guarantee, and there are (non-triangulated) origami examples where second-order solutions do not correspond to points lying in the true configuration space \cite{reciprocalarea}. Nonetheless, the displacements allowed here can only change the stretching energy to at most sixth-order, so we will mostly ignore this issue in what follows and simply refer to our second-order approximations as configurations. Because our analysis is nonlinear, we will end up relying on several more such hypotheses which we have chosen to deal with post hoc by checking that they are satisfied in our numerics, rather than seeking a rigorous proof here.

\subsection{Wheel stresses, Gaussian curvature and single-vertex origami}
\label{subsec:singlevertex}

\begin{figure}
\includegraphics[width=3.5in]{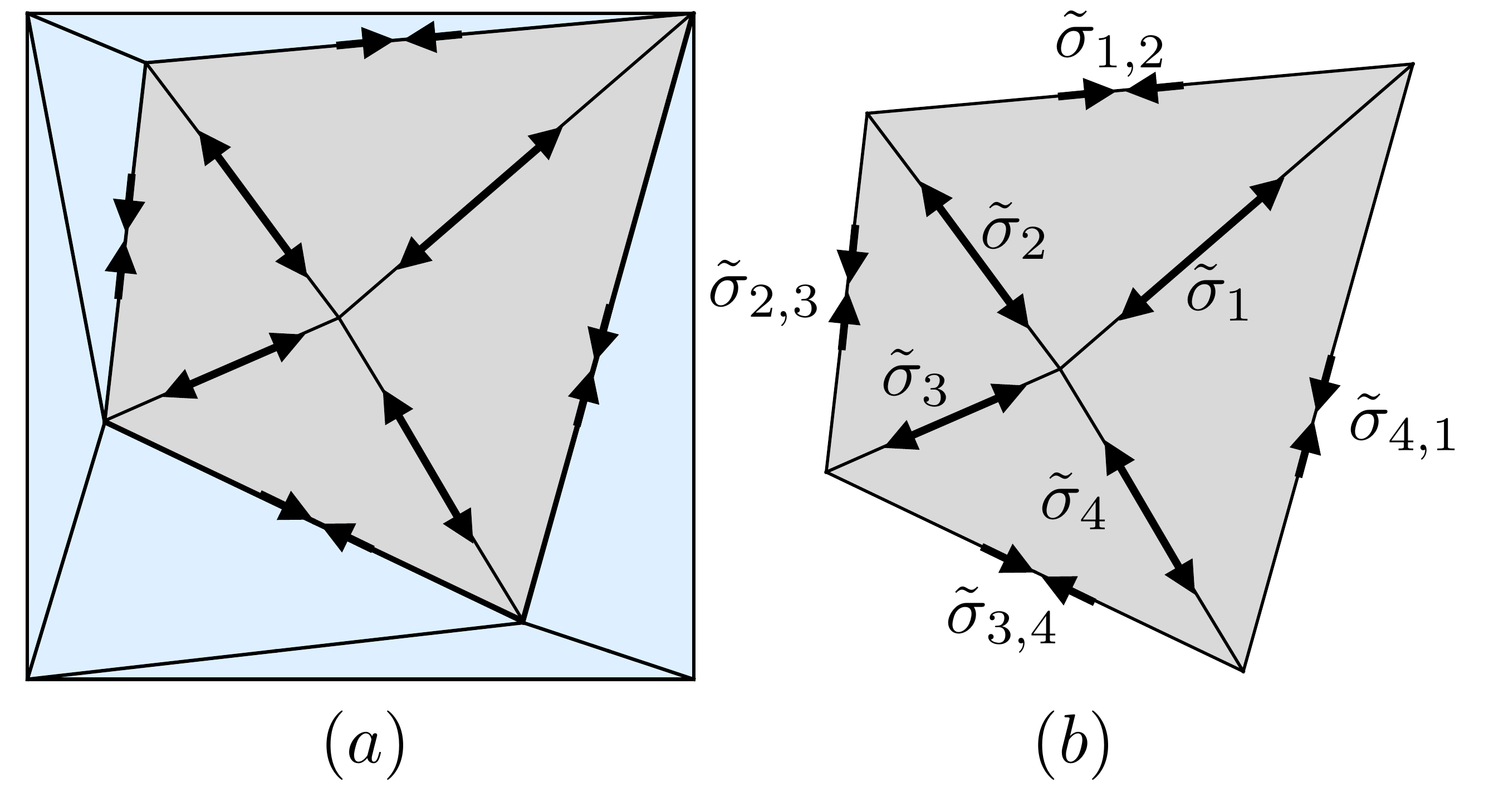}
\caption{\label{fig:wheel} (a) Extracting a single internal vertex from a larger origami structure. (b) The ``wheel stress'' for each single vertex, represented as either extensional or compressional arrows, is also present as a self stress in the larger origami structure of (a). Indeed, such wheel stresses form a basis for the space of self stresses of a generic unfolded triangulated origami, thus the dimensionality of that space is equal to the number of internal vertices. The labels in (b) are associated with Eqs.\ (\ref{wheelstress}).}
\end{figure}

Our first observation is that an unfolded origami structure with $V_i$ internal vertices has at least $V_i$ self stresses. To construct them, we first isolate the faces around each internal vertex and consider the mechanics of the isolated vertex stars apart from the rest of the origami structure (Fig.\ \ref{fig:wheel}a). Those faces and their edges make a spoked wheel of folds emerging from a single internal vertex and meeting the vertices of a polygon. If there are $N$ spokes, the $2N+2$ in-plane positions of the vertices are subject to $2N$ constraints. Since there are three planar Euclidean motions, there must be generically one self stress (Fig.\ \ref{fig:wheel}b). This {\em wheel stress} is preserved if we embed it into the larger structure by setting the remaining components of $\sigma_{(n,m)}$ to zero.

Interestingly, the second-order constraints arising from using the wheel stresses in Eq.~(\ref{eq:compat}) also have a natural geometric interpretation in terms of the Gaussian curvature at each internal vertex, measured by the sum of the angles between adjacent folds around it. To see this, note that we can also generate constraints by enforcing the condition that the Gaussian curvature remains zero at each internal vertex after deformation. We label the folds around each internal vertex with an index $I$ which we take modulo the number of folds meeting at the vertex. Then let $\alpha_{I,I+1}$ be the planar angle between folds $I,I+1$ and let $\psi_I$ be the angle the $I^{th}$ fold makes with respect to the $xy-$plane.  Spherical trigonometry yields the constraint
\begin{eqnarray}
0 &=& \sum_I  \left\{ \frac{\psi_I \psi_{I+1}}{\sin \alpha_{I,I+1}} -\frac{1}{2} \cot \alpha_{I,I+1} \left[\psi_I^2 + \psi_{I+1}^2 \right] \right\},
\end{eqnarray}
valid up to quadratic order in the $\psi_I$. 

To lowest order, we have $\psi_{(n,m)} = (h_n-h_m)/|\mathbf{U}_n - \mathbf{U}_m|$ in terms of the height displacements. We thus have an equation for each internal vertex in the form of Eq.\ (\ref{eq:compat}) from which we can read off a candidate self stress $\sigma_{(n,m)}$. If we denote the self stress on the outside edges (on the rim of the wheel) of each vertex by $\tilde{\sigma}_{I,I+1}$ and the self stress on the spokes $I$ by $\tilde{\sigma}_I$ (Fig.\ \ref{fig:wheel}b), then we obtain
\begin{eqnarray}
\tilde{\sigma}_{I,I+1} &=& -\csc \alpha_{I,I+1} \frac{\Delta L_{I,I+1}}{L_I L_{I+1}}\nonumber\\ \label{wheelstress}
\tilde{\sigma}_I &=& L_{I+1}^{-1} \csc \alpha_{I,I+1} + L_{I-1}^{-1} \csc \alpha_{I-1,I}\\
& & - L_{I}^{-1} \left(\cot \alpha_{I,I+1}+\cot \alpha_{I-1,I}\right)\nonumber,
\end{eqnarray}
where $L_I$ is the length of fold $I$ and $\Delta L_{I,I+1}^2 = L_I^2 + L_{I+1}^2 - 2 L_I L_{I+1} \cos \alpha_{I,I+1}$ (Appendix \ref{app:gc}). In Appendix \ref{app:gcwheelstress}, we prove that Eqs.\ (\ref{wheelstress}) do give the coefficients of a self stress and thus that the wheel stress constraint is precisely Gaussian curvature preservation.

\begin{figure}%[b]
\includegraphics[width=3.5in]{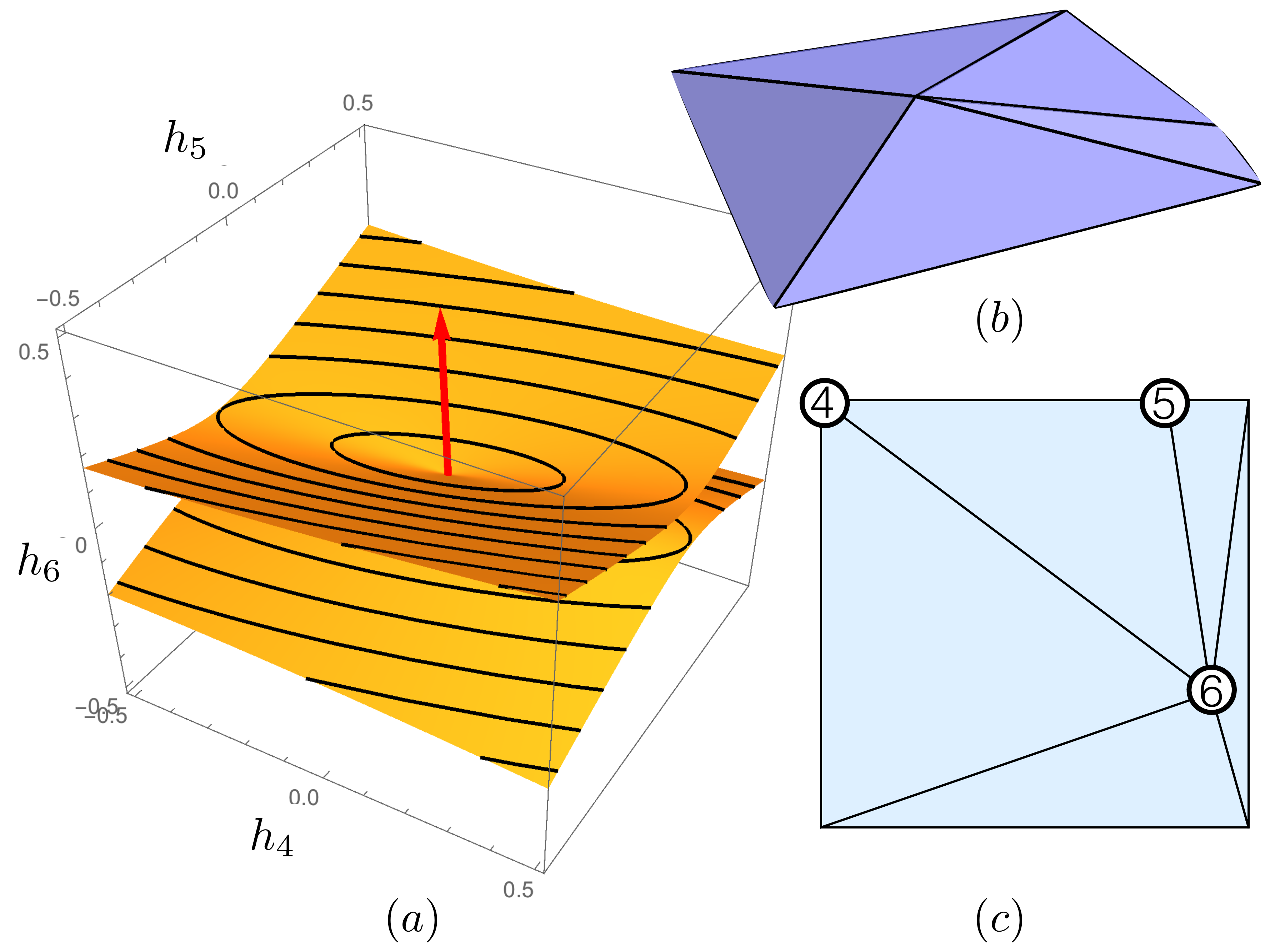}
\caption{\label{fig:conical} (a) The configuration space of a five-fold origami vertex structure, for the crease pattern in (c), where the three unlabeled vertices are pinned to the $xy-$plane. The axes correspond to the vertical displacements of the three vertices labeled $4$--$6$. (b) The eigendeformation of the negative eigenvalue would, on its own, give the vertex positive Gaussian curvature. (c) The planar crease pattern with the three labeled vertices, $4$--$6$.}
\end{figure}

We now discuss the configuration space of single-vertex origami structures. Our analysis here will play a big role in our later treatment of multiple-vertex structures. Let $Q_{nm}$ be the symmetric matrix corresponding to $\bm{\sigma}$ so that the quadratic form in Eq.\ (\ref{eq:compat}) is written in terms of the vector of vertical displacements at vertices $\mathbf{h}$,
\begin{equation}\label{eq:stressmatrix}
\sum_{n,m} Q_{nm} h_n h_m=0.
\end{equation}
If the vertex associated with $Q_{nm}$ has $N$ folds, we find that the $(N+1)\times(N+1)$ matrix $Q_{nm}$ has $N-2$ nonzero eigenvalues, exactly one of which is negative (Ref.~\onlinecite{kapovichmillson} and Appendix \ref{app:quadratic}). This means that a single $N-$fold vertex has $N-2$ first-order motions once translations and rotations have been removed. The second-order motions are the solutions to Eq.~(\ref{eq:stressmatrix}), which defines a $(N-3)$-dimensional surface in the linear space of first-order motions, namely, the null-cone of this quadratic form. Since $Q_{nm}$ has one negative eigenvalue, its null-cone has two conical components ({\em nappes}) that meet at the unfolded state (Fig.~\ref{fig:conical}a). Topologically, the nappes are cones over $(N-4)$-dimensional spheres.

Let $\mathbf{h}_-$ be the eigenvector corresponding to this negative eigenvalue. Points on the two nappes can be distinguished by the sign of their dot products with $\mathbf{h}_-$ since the plane normal to $\mathbf{h}_-$ separates the nappes. This eigenvector gives a set of displacements that maximizes the change in the Gaussian curvature. Indeed, it can be shown from the formulae given in Appendix \ref{app:quadratic} that the component of largest magnitude in this eigenvector is the displacement at the vertex itself and the neighboring vertices are moved by smaller amounts in the opposite direction. Such a displacement (Fig.~\ref{fig:conical}b) leads to a conical deformation at the vertex. This suggests that the difference between rigid origami configurations in the two nappes is related to whether the vertex is buckled up or down (relative to the upwards normal of the origami sheet). We make this more precise in the rest of this subsection.

\begin{figure}%[b]
\includegraphics[width=3.5in]{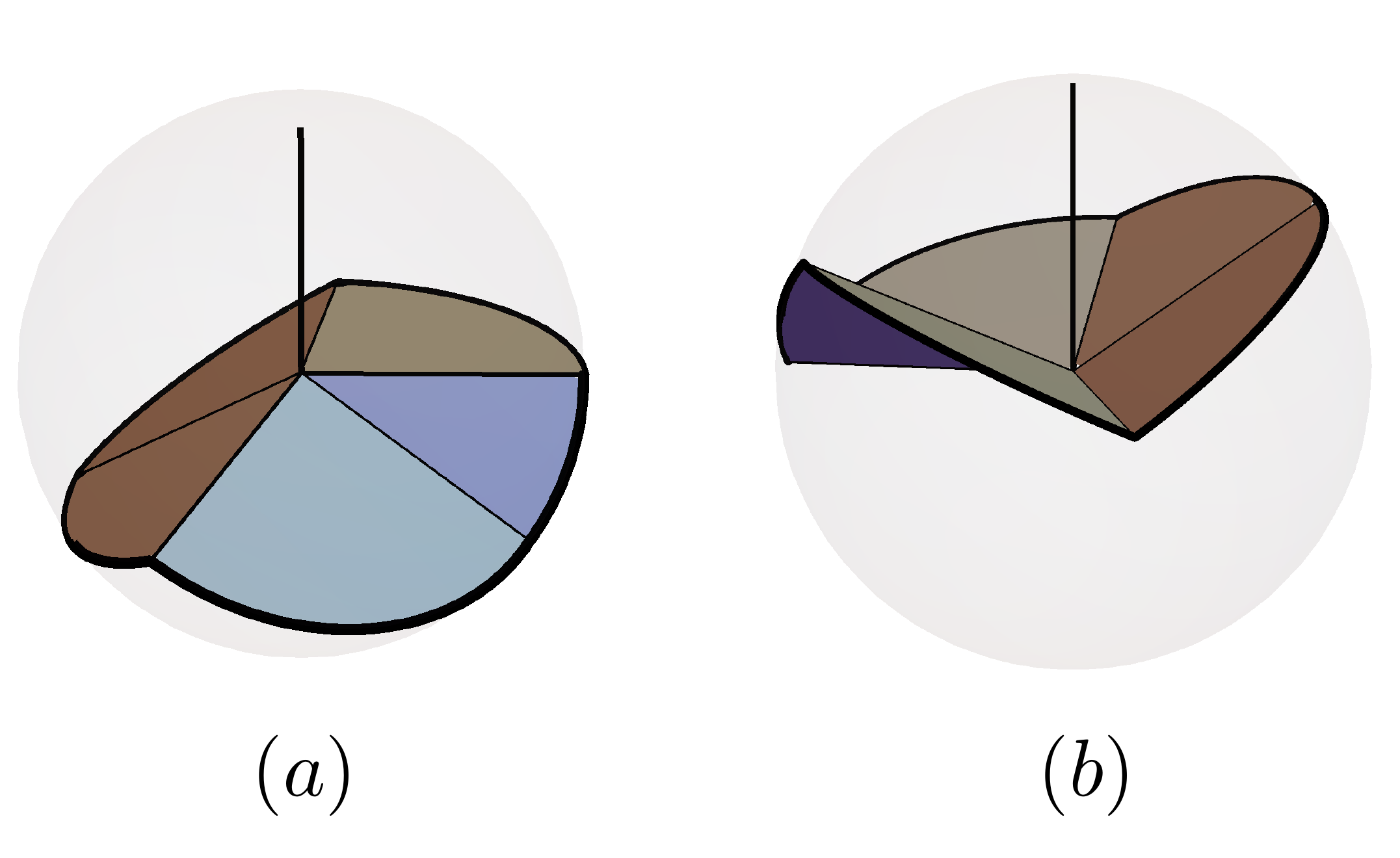}
\caption{\label{fig:popped} (a) A 5-fold origami vertex in a popped up configuration; the vertical line depicts the ``upwards'' normal of the origami surface. (b) The same vertex in a popped down configuration.}
\end{figure}

The {\em trace} of an origami vertex is defined to be the spherical polygon obtained by intersecting the origami with a small sphere centered at the vertex; it is non-self-intersecting for a vertex sufficiently close to being unfolded and thus cuts the sphere into two pieces, corresponding to the upper and lower sides of the origami sheet. Since the trace of a folded origami vertex lies completely in an open hemisphere \cite{streinuwhiteley}, one of those pieces will have area less than $2\pi$. If that piece corresponds to the upper side of the origami, the configuration is called {\em popped down} (as the vertex ``points'' towards the lower side of the sheet) and otherwise it is called {\em popped up} \cite{abel2015rigid} (Fig.~\ref{fig:popped}). Since configurations of these two types meet only in the unfolded state, one of the nappes of the double cone configuration space consists of popped up configurations and the other consists of popped down configurations, so we can use dot products with $\mathbf{h}_-$ to distinguish them computationally. Note that Ref.~\cite{abel2015rigid} give a simpler definition, where popped down (up) vertices are those whose edge vectors are all in the northern (southern) hemisphere. Our definition is a rotation-invariant generalization that is better suited for considering configurations at vertices that are part of larger multiple-vertex origami. 

\subsection{Consequences for multiple-vertex origami and the definition of branches}\label{subsec:count}

Eqs.\ (\ref{eq:compat}) provide a way to count the number of infinitesimal degrees of freedom of an arbitrary triangulated origami structure (subject to the caveats described at the end of Section~\ref{subsec:model}), and provide information on the number of distinct ways of folding a given crease pattern from the unfolded state. 

\begin{figure}%[b]
\includegraphics[width=3in]{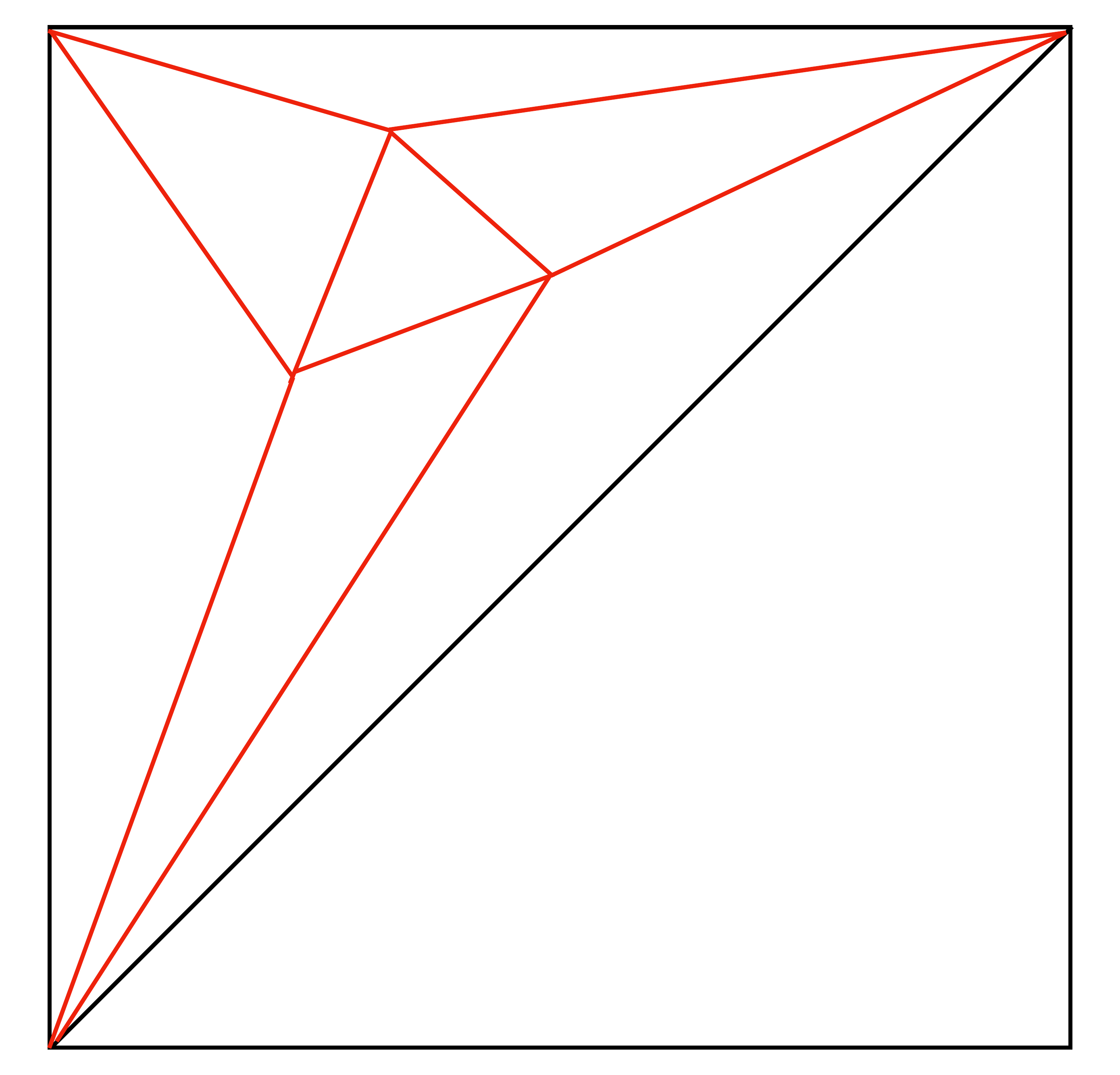}
\caption{\label{fig:triangle} An example of a crease pattern containing a fold joining two external vertices. Such folds do not couple to any others and we will not consider any crease patterns containing these (except for the individual butterflies in Section \ref{sec:butterfly}). This crease pattern also contains a ``triangulated triangle.'' All of the red folds above the upper left diagonal of the square lie within a triangle, hence the three vertices interior to the triangle are rigid even though they each have four folds.}
\end{figure}

Suppose we have an unfolded origami structure with $V_i$ internal vertices and $V_e$ boundary vertices. There are $V_i + V_e$ linear degrees of freedom corresponding to vertical displacements, but $V_i$ quadratic equations constraining them. We are assuming here that all folds are incident to at least one internal vertex (i.e.~we cannot disconnect the crease pattern by cutting along any one fold as in Fig.~\ref{fig:triangle}). Therefore, the number of nontrivial degrees of freedom for a generic unfolded triangulation, should be given by
\begin{equation}\label{eq:Dcount}
D = V_e - 3.
\end{equation}
The term $3$ arises from removing the three remaining out-of-plane Euclidean motions (this can be done by e.g.~pinning the vertices of an arbitrary triangle to the $xy-$plane). Eq.\ (\ref{eq:Dcount}) recovers the count for the degrees of freedom for a generic (folded) triangulated origami derived from a linear analysis \cite{chen2016topological}. The unfolded configuration admits $V_i+V_e-3=D+V_i$ nontrivial linear motions, so this linear analysis fails, though we see that counting quadratic constraints as we do here leads to the expected number $D$. Geometrically, the unfolded configuration is a singular point of the configuration space, where the dimension of the tangent space exceeds the dimension at other nearby points.

More precisely, $D$ should be the local dimension of the configuration space at nonsingular points. However, since the constraint equations are nonlinear, our derivation of Eq.\ (\ref{eq:Dcount}) is not rigorous. To give a proof, we would also need to show that the hypersurfaces arising from Eq.\ (\ref{eq:compat}) for each of the $V_i$ internal vertices intersect transversely (\cite{harris}, Example 11.8). This should be true generically, and in all of the numerical examples considered in this paper this is indeed the case.

Before turning to our discussion of distinct folding pathways and branches, we consider first the case $V_e=3$. Such crease patterns are {\em triangulated triangles} (see the upper left triangle and its red interior folds in Fig.\ \ref{fig:triangle}). Triangulated triangles are equivalent to planar projections (Schlegel diagrams) of triangulations of spheres, \textit{e.g.}~a degree-3 vertex lies in the center of a projected tetrahedron. Our count $D=0$ suggests that these should have 0 degrees of freedom, i.e. that they should be rigid. Gluck proved that generic triangulated spheres are rigid \cite{gluck}. In our case, the triangulated triangles are flat and thus nongeneric, but Connelly proved that these are rigid at second-order in 3D as well \cite{connellycabled}. Thus, in this case, the configuration space is simply a point at the unfolded state (with multiplicity, as we will see later).

To better understand the neighborhood of the unfolded state and to define the notion of ``branches'' when $D\geq1$, we consider the solutions of Eqs.\ (\ref{eq:compat}), which are sets of vectors $\mathbf{h}$ in $\mathbb{R}^{V_i+D}$. Since the quadratic equations are all homogeneous in the vertex heights, any vector $\lambda \mathbf{h}$ solves Eqs.\ (\ref{eq:compat}) if $\mathbf{h}$ does, for any real number $\lambda$. 

Therefore, we are led to consider solutions of Eqs.\ (\ref{eq:compat}) in {\em projective space} $\mathbb{RP}^{V_i + D - 1}$, where a height vector $h$ is identified with $\lambda h$ for any nonzero real number $\lambda$. Let $B$ be the number of connected components of the solution set (counted with multiplicity) in $\mathbb{RP}^{V_i+D-1}$ and let us assume that $B>0$. Roughly speaking each of these connected components is generically a $(D-1)$-dimensional component of the intersection of a small sphere centered at the unfolded state with the origami configuration space, where components related by the $z$-reflection symmetry $h \mapsto -h$ are identified. In this paper we will consider mostly the case where $D=1$ where these components are simply points. Back in the space of first-order deformations $\mathbb{R}^{V_i+D}$, each of these components induces a double cone over some reflection-related pair of components on this sphere, and all of the origins of these cones (singular when $D>1$) intersect at the unfolded state. We will refer to these $B$ cones as {\em branches}. When $D=1$, these cones are simply lines.

Note that by our discussion in Section \ref{subsec:singlevertex}, single-vertex origami structures have a single branch. For the example in Fig.\ \ref{fig:conical}a, where $V_i=1,D=2$, the branch is a double cone over a closed curve in $\mathbb{RP}^2$.

\section{Numerical results}\label{sec:numerical}

\subsection{Counting branches}
\label{subsec:countingbranches}

We now discuss the case of $D=1$ in more concrete terms. In that case, each of the branches corresponds to a curve that passes through the unfolded state, as in Fig.\ \ref{fig:schematic}. To understand the typical number of branches $B$, we consider a class of random, one-degree-of-freedom origami, generalizing the example of Fig.\ \ref{fig:schematic}. We generated random origami structures by computing a Delaunay triangulation on the point set consisting of $V_i$ random points uniformly distributed within a square, together with the corners of the square. We omitted configurations having 3-fold vertices since such vertices are always rigid. We also found no triangulations with edges connecting opposite corners of the square (such as in Fig.\ \ref{fig:triangle}). Such edges would break the square into two rigid triangulated triangles. 

Since there is a vertex at each corner of the square, $V_e = 4$ and so Eq.\ (\ref{eq:Dcount}) yields $D=1$, no matter how the interior vertices and edges are arranged. We fix the overall position and orientation by setting the height of the vertices of one triangle to zero. To remove the scaling symmetry from the homogeneous system coming from Eqs.\ (\ref{eq:compat}), we add a normalizing equation $\sum_j h_j^2=1$, resulting in $V_i+1$ quadratic polynomials that must be simultaneously solved in terms of $V_i+1$ vertex heights. We solved these systems in \textit{Mathematica 11}, which uses a homotopy continuation algorithm for numerical root finding of polynomial systems \cite{morgan1987computing}. Note that the solutions of these polynomial systems come in pairs related by multiplication by -1 corresponding to $z$-reflection symmetry (as discussed above, each branch is a line through the origin and these intersect the normalizing unit sphere twice), so the number of branches $B$ is {\em half} the number of real solutions.

\begin{table}[]
\centering
\caption{Summary of random triangulation computations: $V_i$ is the number of internal vertices of the triangulation, the column ``systems with $V_t>0$'' gives the number of triangulations found that include triangulated triangles (in all cases these were flat octahedra), the column ``MV duplicates'' gives the number of triangulations that included at least one pair of branches with the same mountain and valley fold assignments. In all cases, the number of branches with multiplicity was $2^{V_i}$, the number of distinct branches was $2^{V_i-V_t}$ and the branches were in 2 to 1 correspondence with their vertex sign patterns (defined in Section \ref{subsec:vsp}).}
\label{tab:summary}
\begin{tabular}{l|lllll}
$V_i$ & \begin{tabular}[c]{@{}l@{}}triangulations \\ generated\end{tabular}
& \multicolumn{1}{c}{\begin{tabular}[c]{@{}c@{}}precision\\ used\end{tabular}} & \multicolumn{1}{c}{\begin{tabular}[c]{@{}c@{}}systems\\with\\$V_t > 0$\end{tabular}} & \begin{tabular}[c]{@{}l@{}}MV\\ duplicates\end{tabular} \\ \hline
2     & 100                                                                 & 500                                                                                & 0                                                                                     & 0                                                               \\
3     & 5000                                                                & 690                                                                                & 0                                                                                     & 0                                                               \\
4     & 1000                                                                & 690                                                                                & 2                                                                                     & 0                                                               \\
5     & 1000                                                                & 690                                                                                & 1                                                                                     & 17                                                               \\
6     & 1000                                                                & 690                                                                                & 1                                                                                     & 28                                                               \\
7     & 300                                                                 & 690                                                                                & 0                                                                                     & 17                                                               \\
8     & 50                                                                 & 690                                                                                & 0                                                                                     & 8                                                              
\end{tabular}
\end{table}

We generated several thousand random triangulated origami squares with $V_i=2$ to 8 and computed and analyzed their branches (Table \ref{tab:summary}). Accurate solutions seem to require very high precision arithmetic, especially as $V_i$ becomes larger; to ensure good results we used up to $690$ digits of precision and verified the resulting solutions, which allowed us to find solutions up to $V_i = 8$. 

Using these second-order branches, we numerically computed approximate configurations (solutions to the length equations in Eq.\ (\ref{eq:length})) to create our figures and movies by using the solutions to perturb the unfolded configuration and minimizing a stretching energy (sum of squared differences of edge lengths to the lengths in the unfolded state) on the coordinates until it was zero to high accuracy. 

We now give a little bit of background on systems of polynomial equations to give context for our main results. In general, there is little one can say about the simultaneous roots of a completely arbitrary system of polynomials, especially if one is interested in real roots. One result, known as B\'ezout's theorem, states that if the solutions to a system of polynomial equations are isolated points, then the number of complex solutions, counted with multiplicity and including points ``at infinity'', is equal to the product of the degrees of the equations (\cite{harris}, Lecture 18). For our systems of $V_i+1$ quadratic polynomials in $V_i+1$ height variables, this yields $2^{V_i+1}$ roots. However, we are only interested in real and finite roots. Our systems have the property that all coefficients are real, but this merely guarantees that nonreal roots come in complex-conjugate pairs; there might still be no real roots. 

Now we state the first of our main numerical findings: with only a few exceptions we will discuss shortly, all the solutions of our system are distinct and isolated and amazingly, all $2^{V_i+1}$ of them are real! Since the branches correspond to $\pm$ pairs of solutions, we therefore have $B=2^{V_i}$.

As mentioned, not all crease patterns lead to $2^{V_i}$ {\em distinct} branches. We found a very small number of systems ($2$ with $V_i=4$, $1$ with $V_i=5$ and $1$ with $V_i=6$, see Table \ref{tab:summary}) with fewer branches; in these cases all branches came with some multiplicity. In these cases, we were able to identify triangulated triangles within the crease pattern. Let $V_t$ be the number of vertices in the interior of all such triangulated triangles (if we had not excluded crease patterns with degree-3 vertices from our computations, we would count these in $V_t$). These vertices must remain unfolded in all branches of configurations, and it follows that their height variables do not contribute to distinct roots of the quadratic equations, but rather only give rise to multiplicity.  We found that in such cases, the distinct roots of the quadratic equations had multiplicity $2^{V_t}$. Thus we may account for this effect with the (conjectural) formula
\begin{equation}
B_{distinct}=2^{V_i-V_t}.
\end{equation}
Let us call the $V_i-V_t$ internal vertices which are not interior to triangulated triangles {\em foldable vertices}. Note that the heights of non-foldable vertices are determined by the foldable ones by the linear condition that each triangulated triangle is planar, so in the space of linear motions $\mathbb{R}^{V_i+1}$ the branches must lie in a lower-dimensional $\mathbb{R}^{V_i-V_t+1}$.

We now briefly discuss the pattern of mountain and valley folds in the branches.
Following a short Euler characteristic argument, the number of folds (internal edges) for a triangulated square is $3 (V_i-V_t+ 1)$, so the potential number of distinct mountain-valley (MV) assignments up to a global sign change is $2^{3 (V_i-V_t+1)}/2$. Na\"ively, the fraction of these MV assignments that we should expect to find among the branches is at most $(1/4)^{V_i-V_t-1}$, which approaches zero exponentially in the number of vertices. However, there are combinatorial consistency constraints on the MV patterns, along the lines of those derived in Ref.~\cite{abel2015rigid} for single-vertex origami, so this is certainly an overestimate. We did not attempt to work out these consistency conditions, but as some evidence that they play a role, in our computer-generated examples with $V_i=5,6,7,8$ we found an increasing number of crease patterns where multiple branches have coinciding MV assignments (Table \ref{tab:summary} and Section \ref{subsec:nonunique}). This phenomenon is well-known in the origami community \cite{tomhullpersonal} and an illuminating example is the 6-fold origami vertex with alternating mountain and valley folds (MVMVMV in cyclic order around the vertex) called the ``waterbomb structure'' \cite{brunck2016,tachi2016self}. Since the distribution of mountain and valley folds do not distinguish different branches in configuration space, and further, since it is hard to guess which MV assignments are allowed, the question remains, is there anything that does distinguish those branches from each other?

\subsection{Vertex sign patterns}
\label{subsec:vsp}

Given a folded configuration of an origami square with multiple internal vertices, we can ask whether each vertex is popped up or down. This data is encoded as {\em vertex sign patterns}, assignments of $+1$ or $-1$, to the foldable vertices if they are popped up or down, respectively. (Since the $V_t$ vertices lying within triangulated triangles are always unfolded, we could extend the vertex sign pattern to these vertices by assigning them the value 0). As there are therefore $2^{V_i-V_t}$ choices of vertex sign patterns, it is natural to hope that there is a one-to-one correspondence with the branches. However, branches consist of pairs of solutions related by the $z$-reflection symmetry, so we must identify vertex sign patterns related by a global sign change and we are left with only $2^{V_i-V_t-1}$ equivalence classes. Note that in a single 4-fold vertex origami, we have only one vertex sign pattern up to sign and two branches, so instead the best we can hope for is a one-to-one correspondence between {\em pairs} of branches and sign-related pairs of vertex sign patterns.

We check this correspondence computationally as follows: we first determine the eigenvector with negative eigenvalue of $Q_{nm}$ for each internal vertex $n$, $\mathbf{e}_n$ (shown as an arrow in Fig.\ \ref{fig:conical}). Here we ensure that $+\mathbf{e}_n$ corresponds to the popping up deformation and we extend the vector with zero components so that its dimension is the same as that of $\mathbf{h}$. Then the vertex sign pattern is defined by $\sigma_n = \textrm{sgn} [\mathbf{e}_n \cdot \mathbf{h}]$. Both $\mathbf{h}$ and $-\mathbf{h}$ correspond to the same branch, so we associate to each branch a pair of vertex sign patterns related by a global sign change. Our second main numerical finding is the following: remarkably, in {\em all} of our computed examples (summarized in Table \ref{tab:summary}), there are exactly two branches with each such pair of sign patterns, in agreement with our guess above!

\begin{figure}%[b]
\includegraphics[width=3.5in]{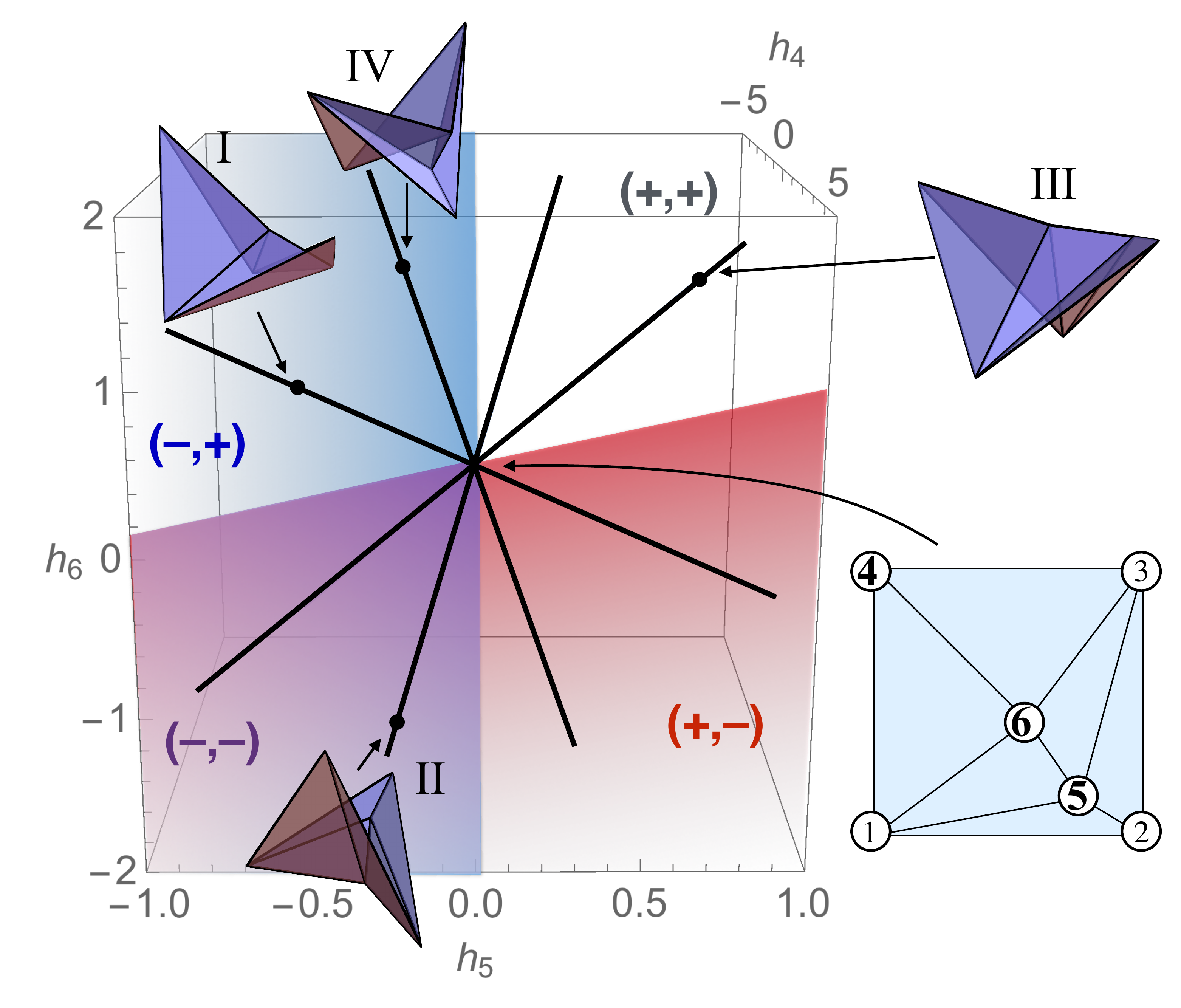}
\caption{\label{fig:chambers} A view of the configuration space shown in Fig.~\ref{fig:schematic} embedded in the space of height coordinates, rather than fold-angle coordinates, with the four folded configurations from branches I-IV superimposed. Only the second-order approximations to the branches are plotted here; since at this order they consist of intersecting lines, the height units are arbitrary. The upper / lower sides of the representative configurations have been colored blue / red, respectively, to make the popping at each vertex more evident. The space is divided into four wedge-shaped 3D chambers (unshaded, red, purple, blue) corresponding to the four possible pairs ($(+,+), (+,-), (-,-), (-,+)$) of signs of the two quantities $s_5=\mathbf{e}_5\cdot \mathbf{h}$ and $s_6=\mathbf{e}_6 \cdot\mathbf{h}$, where $\mathbf{e}_5,\mathbf{e}_6$ (not drawn) are the negative-eigenvalue eigenvectors corresponding to the internal vertices 5,6, respectively. Thus for example, the unshaded chamber $(+,+)$ consists of all height deformations $\mathbf{h}$ where vertices 5 and 6 are both popped up. The viewpoint has been chosen carefully to be ``edge-on'' to the planes $s_5=0$ and $s_6=0$. From this angle, the two planes are projected onto the lines separating the colored regions in the figure, and the line $s_5=s_6=0$ along which these planes intersect becomes the line passing through the origin that is normal to the plane of the figure.}
\end{figure}

If we look at the branches as lines intersecting in the singular unfolded state in the second-order configuration space $\mathbb{R}^{V_i-V_t+1}$, each of the $\mathbf{e}_n$ defines a hyperplane separating the popped-up configurations at $n$ from those popped-down there (Fig.~\ref{fig:chambers}). The set of all such planes arising from the $V_i-V_t$ foldable vertices divides $\mathbb{R}^{V_i-V_t+1}$ into $2^{V_i-V_t}$ chambers, each labeled by a different vertex sign pattern. Each of these chambers is topologically the product of an orthant of $\mathbb{R}^{V_i-V_t}$ with a real line. For instance in Fig.~\ref{fig:chambers}, each chamber is topologically the product of a quarter-plane with a line, a ``wedge-shaped'' region of 3D space (the projection has been specially chosen so that these chambers are seen ``edge-on'', otherwise the dividing planes obstruct the view). In these terms, our observation is that the $2^{V_i-V_t+1}$ distinct rays of the branches always seem to be distributed so that two rays lie in each of these chambers, implying that the solutions to the coupled system of nonlinear equations Eq.\ (\ref{eq:compat}) are controlled to some extent by what happens at each vertex. Again, from the perspective of random solutions to real quadratic equations, one might have expected some of these branches to be complex conjugate pairs and that the real branches would be distributed much more unevenly in the chambers. Simple tests with random perturbations of the coefficients of our equations (so that they no longer come from realizable crease patterns) confirm this expectation -- after perturbing, the solutions of most systems have many complex-conjugate pairs and real solutions are not equidistributed.

\subsection{Branches with non-unique Mountain-Valley assignments}
\label{subsec:nonunique}

We now have two pieces of combinatorial data associated to each branch: first, the well-known assignment of which folds are mountain and which are valleys (MV assignment) and second, the vertex sign pattern (modulo sign). For most of the origami crease patterns we computed (Table \ref{tab:summary}), the branches all have different MV assignments. However, this is not always the case, as mentioned at the end of Section \ref{subsec:countingbranches}. When pairs of branches have identical MV assignments, they can usually be distinguished by the vertex sign patterns; in particular there is usually exactly one vertex sign that differs. However, interestingly, we did find a number of examples where the two branches with the MV assignment also have the same vertex sign pattern. In the rest of this section we will show examples of these occurrences.

\begin{figure}%[b]
\includegraphics[width=3.5in]{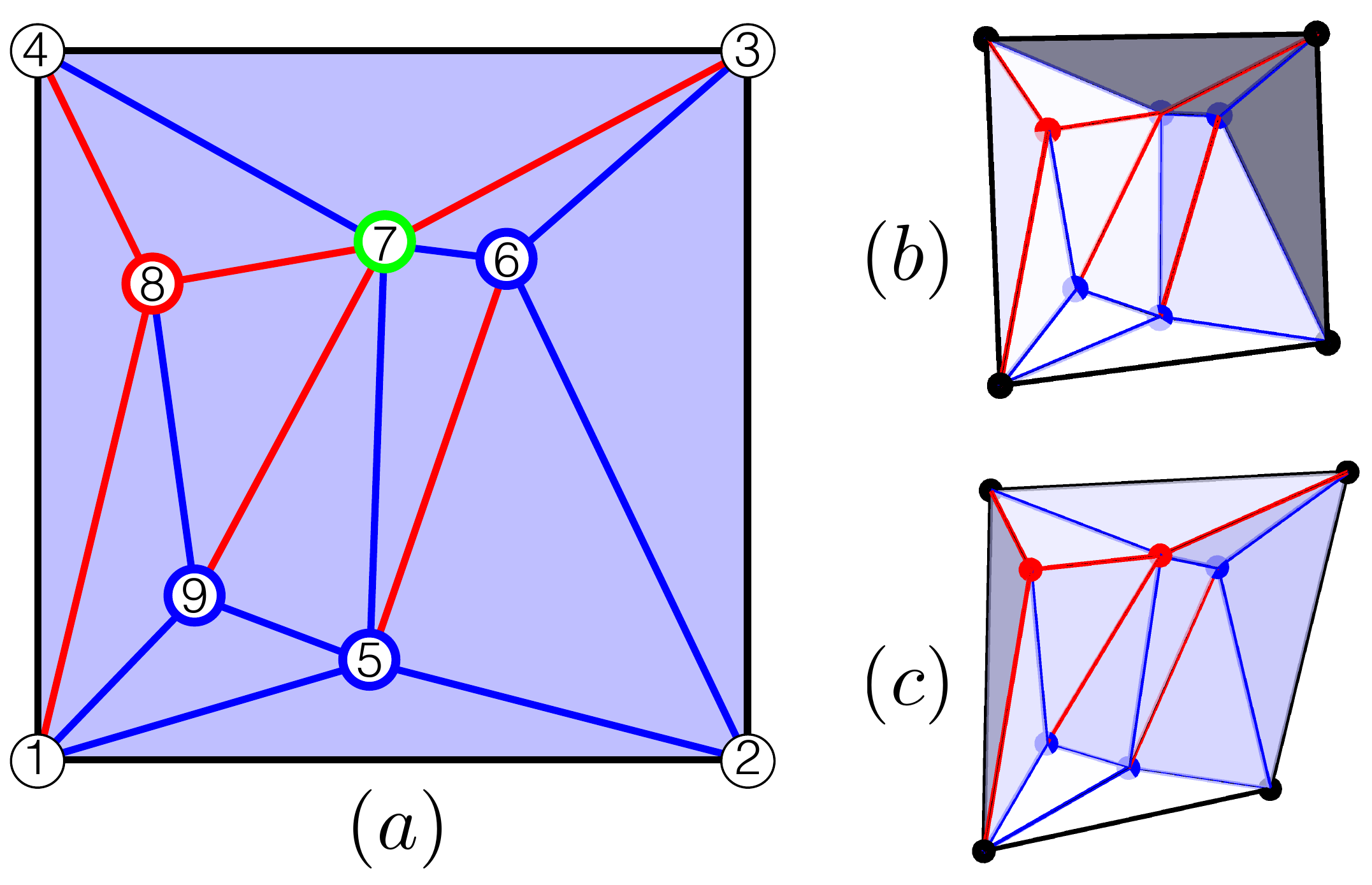}
\caption{\label{fig:mvexample1} (a) Crease pattern of a triangulated square with $V_i=5$ with a pair of branches where the MV assignments to the edges are identical. Edges colored red and blue represent mountain and valley folds, respectively. Vertices with red and blue outlines represent vertices that are popped up and popped down in both branches, respectively. The state of the green vertex (labeled 7) distinguishes the two branches; i.e.~it is popped up in one branch and popped down in the other. (b), (c) give images of folded configurations on the two branches. Supplementary Movies 5-6 show animations of the folding motions along these branches from the unfolded state.}
\end{figure}

In Fig.~\ref{fig:mvexample1} we show a typical origami crease pattern from our data exhibiting two noncongruent branches with coincident MV assignments. The branches in this example (and most of the other ``MV-coincident pairs'' we found) can be distinguished by the popping state of exactly one vertex. It follows from Corollary 1 of Ref.~\cite{abel2015rigid}, that a vertex which can both pop up and pop down must have degree at least 6, and indeed, must contain both a mountain ``bird's foot'' and a valley bird's foot as subsets of the folds around the vertex. Here, a bird's foot is a sequence of four not-necessarily adjacent folds $c_1,c_2,c_3,c_4$ in counter-clockwise order around the vertex such that the angles between $c_1,c_2,c_3$ are between 0 and $\pi$ and $c_1,c_2,c_3$ have the same sign (all mountains or all valleys) and $c_4$ has the opposite sign. Note that the waterbomb vertex contains both a mountain bird's foot and a valley bird's foot.

\begin{figure}%[b]
\includegraphics[width=3.5in]{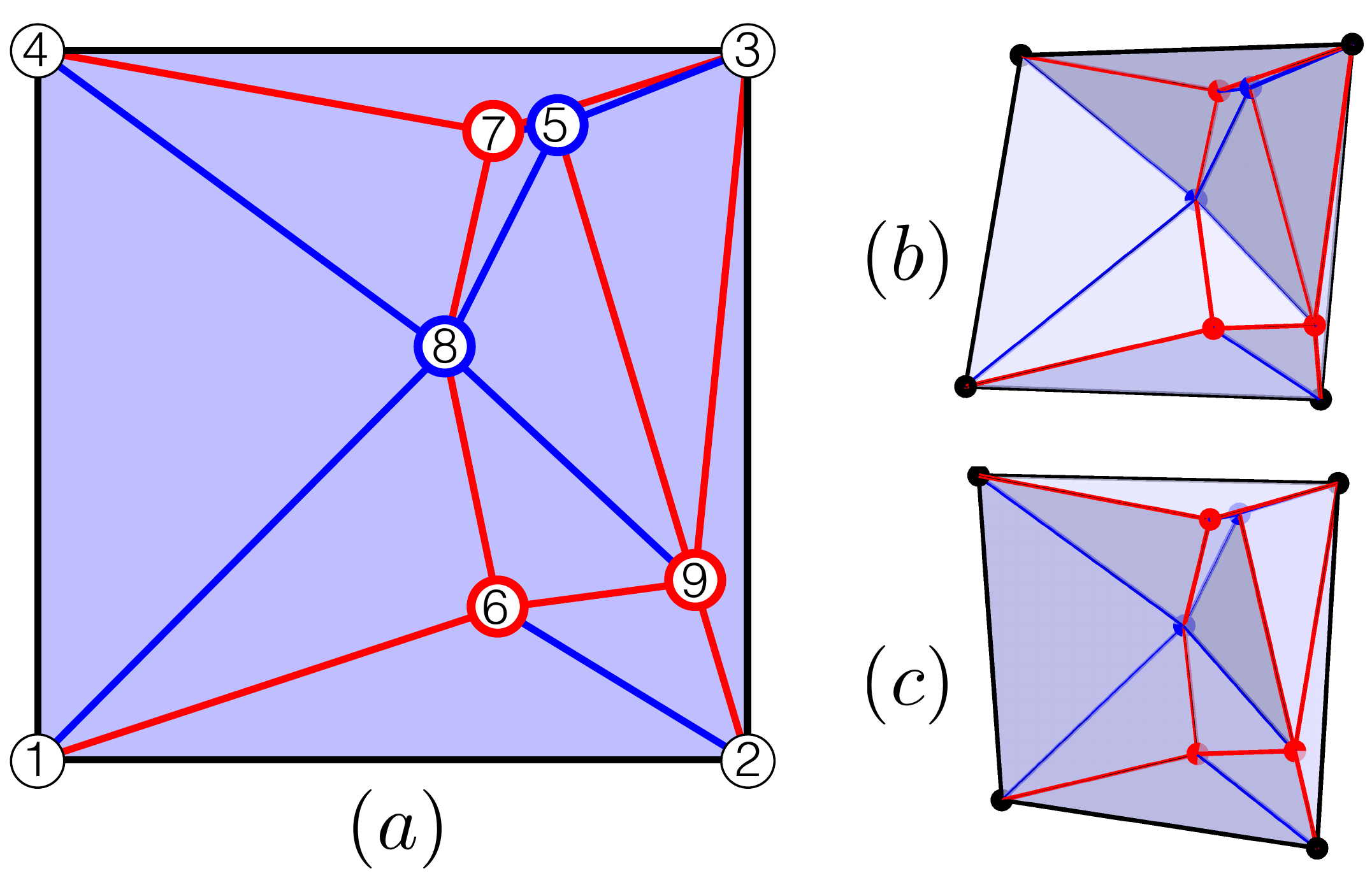}
\caption{\label{fig:mvexample2} (a) Crease pattern of a triangulated square with $V_i=5$ and a pair of branches with coinciding MV assignments and vertex sign patterns, with coloring as in Fig.~\ref{fig:mvexample1}. There is a short fold between vertices 5 and 7 which is a valley fold. In this pair of examples, all vertices are popped the same way in both branches. (b), (c) give images of folded configurations on the two branches. Supplementary Movies 7-8 show animations of the folding motions along these branches from the unfolded state.}
\end{figure}

More interestingly, we found a few examples (1 configuration with $V_i=5$, 2 with $V_i=6$ and 1 with $V_i=7$, out of the configurations computed for Table~\ref{tab:summary}) where MV-coincident branches also had the same vertex sign patterns. The example with $V_i=5$ is shown in Fig.~\ref{fig:mvexample2}. 

In all the examples we computed where branches had coincident MV assignments, they either had exactly one high-degree vertex whose popping state distinguished the branches, or none. We do not dare venture to guess about the relative frequency of such examples as $V_i$ gets large.

\section{H1 triangulations and butterflies}\label{sec:butterfly}

\begin{figure}
\includegraphics[width=3.5in]{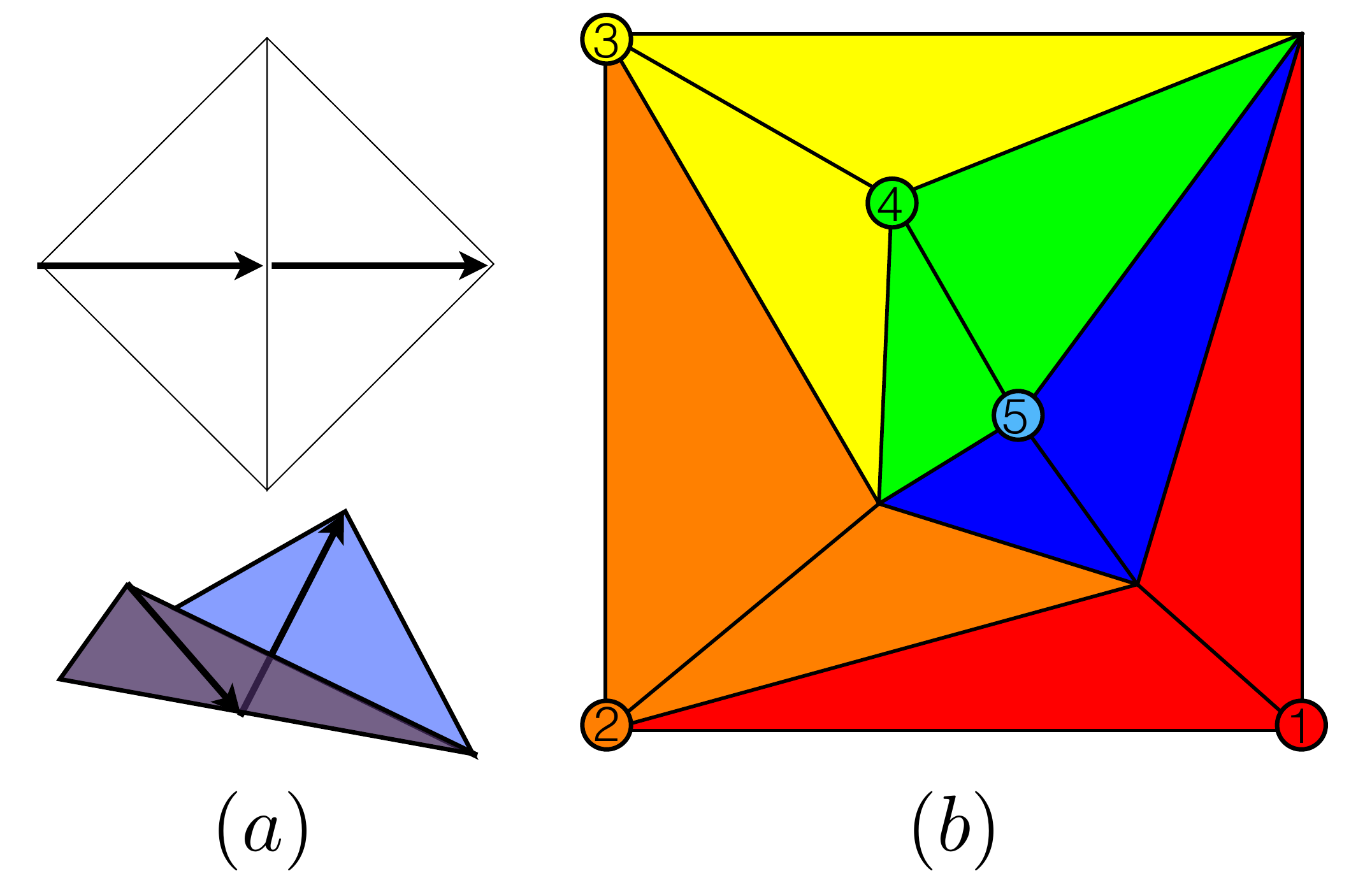}
\caption{\label{fig:peelable} (color) (a) An illustration of the distance-maximization property of the unfolded state for a ``butterfly''. (b) An H1 triangulated origami with $V_i=4$. One possible reduction sequence is depicted, with butterflies and their associated boundary degree-3 vertices colored and labeled in order: (1) red, (2) orange, (3) yellow, (4) green, and ending on (5) the blue seed.}
\end{figure}

In this section we describe a class of $D=1$ triangulations whose configuration spaces are particularly easy to analyze.  We will show that near the unfolded state, their configuration spaces consist of $2^{V_i}$ intersecting 1D branches. In contrast to most of the rest of our results, we will be discussing actual configurations here, not just second-order approximations. 

Henneberg moves \cite{graverservatiusservatius,taywhiteley1985,nixonross2012} are rigidity-preserving transformations of graphs that are useful in rigidity theory. We will only consider the Henneberg-1 move, which in three dimensions is just the addition of a new vertex and attaching it to the original graph with three new edges. In particular, we are interested in triangulations that can be built by repeatedly {\em adding} such degree-3 vertices to the boundary. To be more precise, suppose we have some triangulation $T$. First choose three vertices $v_1,v_2,v_3$ on the boundary of $T$ that are adjacent in the cyclic ordering. Then add a new vertex $w$ and join it with new edges to $v_1,v_2,v_3$. We will say that $T$ and the resulting triangulation $T'$ are related by an {\em H1 move} and will not refer to general Henneberg-1 moves any more. See Fig.~\ref{fig:peelable}(b), where the vertex labeled (1) has been added by an H1 move to the rest of the triangulation; note how (1) and the three edges incident to it form a pair of adjacent red triangles on the boundary of the triangulation. We will require that $v_2$ has degree at least 3, as otherwise, it would become a degree-3 vertex in the interior of $T'$, and thus be non-foldable.

The key property of H1 moves is that given a configuration of $T$ where the faces around $v_2$ are in a folded configuration, there are always two distinct folded configurations of $T'$. To see this, we consider the motion of butterflies.  A {\em butterfly} is the 1-degree-of-freedom rigid origami consisting of two triangular faces joined by a shared fold (Fig.\ \ref{fig:peelable}a). Its configuration space is topologically a circle, parametrized by the dihedral angle at the shared fold. Consider the distance $d$ between the two non-shared vertices of the butterfly. As the dihedral angle varies from 0 to $2\pi$, the distance $d$ increases monotonically from its minimum value $d_{min}$ in the flat folded state, to its maximum value $d_{max}$ when the butterfly is flat and unfolded, and then decreases monotonically again to $d_{min}$. Thus for each value $d_{min}<d<d_{max}$, there are two distinct configurations of the butterfly that are related by reflection. Now the claim follows since an H1 move can be viewed as gluing a butterfly at $v_1,v_2,v_3$. This is essentially the 3D version of a construction for graphs embedded in the plane described in Ref.~\cite{Borcea2004}. Note that when the faces around $v_2$ are unfolded, then $d$ is maximized and we can only attach the butterfly in its unfolded state.

The basic idea in the rest of this section will be that triangulations that are reducible by reverse H1 moves to a seed triangulation with a simple configuration space can also be understood easily. Unfortunately, it seems that very few triangulations are reducible at all by a reverse H1 move. We generated 20000 triangulations with $V_i=3$ to 8 by the method described in Section \ref{sec:numerical} and found that an increasing fraction of triangulations had no degree-3 vertices at the boundary (starting from 0\% at $V_i=3$ to 88.5\% at $V_i=8$). Furthermore, the number of triangulations in this data set that can be reduced further with more reverse H1 moves appears to decay exponentially. 

Nonetheless, we now narrow our focus to triangulations that can be constructed from a sequence of H1 moves from a butterfly (the {\em seed}).  We will call these {\em H1 triangulations} (Fig.~\ref{fig:peelable}b). In line with our observations in the last paragraph, we found that the fraction of H1 triangulations decreased roughly exponentially from 1 at $V_i=3$ to 0.0077 at $V_i=8$. (It's not hard to check that under our restriction of no interior degree-3 vertices, all $V_i=1,2,3$ triangulations are H1). 

Note that each reverse H1 move results in the deletion of one boundary vertex and the conversion of one internal vertex to a boundary vertex. This means that a H1 triangulation with $V_i$ internal vertices can be decomposed into a sequence of $V_i$ butterflies. Also, since the seed butterflies always have four boundary vertices, every H1 triangulation also has four boundary vertices. We will assume in this section that the triangulations have no interior degree-3 vertices and also that there are no collinear folds meeting at vertices (there are no ``crosses'' in the terminology of Ref.\ \cite{abel2015rigid}), as these nongeneric collinearities can rigidify folds in a flat state.

Let us now discuss the configuration space of H1 triangulations near the unfolded state. As discussed above, the seed butterfly has a configuration space that is a circle. Now consider the butterfly that is attached by the first H1 move. We showed that there are two distinct ways of attaching it when the seed is nonflat, and one way of attaching it when the seed is unfolded. We can always choose a small enough interval around the unfolded state in the configuration space of the seed where the dihedral angle of the new butterfly never reaches $\pi$. This implies that a neighborhood of the unfolded state of the configuration space of the two butterflies together consists of two intersecting lines, topologically a letter X, since we have two choices over every nonzero initial dihedral angle, glued together at the unfolded state. Continuing, one sees that for each H1 move, the number of 1D branches doubles, and they all still intersect at the unfolded state. This results in $2^{V_i}$ branches, as desired. 

We have not been able to show that all vertex sign patterns are realized twice, as this seems to require some careful analysis of the configurations of the boundary vertices and when they result in popped up / down states of interior vertices after an H1 move. However, one can generalize the arguments in this section to H1-like triangulations where the seed is not just a butterfly but some other simple triangulation, e.g.~a single-vertex origami. We hope to elaborate on this elsewhere.

\section{Discussion}\label{sec:discussion}

Our results have broad importance in using origami techniques to manufacture shapes from flat substrates. In the standard paradigm for self-folding origami, an initially unfolded sheet is ``programmed'' by setting the equilibrium dihedral angle of each fold to a nonzero value \cite{Silverberg2014,metasheets,tachi2016self,brunck2016}. One illustrative self-folding energy functional takes the form
\begin{equation}
E[x]=\frac{1}{2}\sum_{(n,m)}k_{(n,m)}(\theta_{(n,m)}-\bar\theta_{(n,m)})^2,
\end{equation}
where $\theta_{(n,m)}$ is the folding angle of fold $(n,m)$, $\bar\theta_{(n,m)}$ is the programmed equilibrium value, $k_{(n,m)}$ is an angular spring constant for the fold, and the sum is taken over all folds $(n,m)$. One can visualize geometrically this functional in Fig.\ \ref{fig:schematic} as a generalized squared-distance function from the point $\bar\theta$ corresponding to the programmed folding angles.  Based on the branched structure of configuration space we have found (as exemplified by Fig.\ \ref{fig:schematic}), once an origami structure begins folding along the wrong branch, it is potentially very difficult to return it to the desired branch. 

This possibility motivated Tachi and Hull, in Ref.\ \cite{tachi2016self} to introduce the notion that the bending moments driving a self-folding origami structure should drive the pattern in a direction \textit{perpendicular} to any undesirable branches in configuration space; i.e.\ the gradient of the energy functional at the unfolded state should project onto only one branch. Their condition can be justified from the point of view of an analytic energy landscape. In order to avoid mis-folding, the energy should not decrease along any undesired branches. If we assume that the energy of a structure depends only on the angles of the folds, $E(\theta_1,\theta_2,\cdots,\theta_{N_F})$, then naturally we require that any infinitesimal change in fold angles $\Delta \theta_i$ along an undesired configuration branch satisfy
\begin{equation}
\sum_{i=1}^{N_F} \frac{\partial E}{\partial \theta_i} \cdot \Delta \theta_i \ge 0.
\end{equation}
Since branches are symmetric under $\theta\rightarrow-\theta$, $\partial E/\partial \theta_i$ must then be perpendicular to all undesired branches in the $N_F$-dimensional space of fold angles.

In order for this to be satisfied easily, one could hope that under some circumstances, an origami structure would have very few branches; however our results suggest that in generic triangulated patterns with $D=1$, the number of branches depends only on $V_i$ and not the geometry. If we combine the fact that in a triangulation, the number of internal folds is $N_F= 3V_i+const$ with our finding that the number of branches is $2^{V_i}$, we have thus shown that it is impossible to satisfy the above orthogonality constraint for even modest numbers of internal vertices! Nevertheless, it is an interesting open question whether this condition could be satisifed approximately, \textit{i.e.}~whether, for sufficiently large origami crease patterns, we might have most branches within 99\% of being orthogonal, as this happens for random vectors in high dimensions \cite{caifanjiang}. Some preliminary results have been collected in Appendix \ref{app:highD} which show that in fact branches tend to be less orthogonal in fold-angle space than a uniformly random set of vectors. It remains to be seen how this impacts the programmability of self-folding origami structures.

We also note that our methods can be applied even to non-triangulated origami. In that case, we would first triangulate the origami, then add additional quadratic constraints to rigidify certain folds. The dimension of the configuration space becomes $D = V_e - 4 - E$, where $E$ is the number of rigidified folds in the triangulation that are not in the original pattern. When this formula becomes negative, the constraints in the system become redundant and we instead take $D=0$. We would also like to understand what happens when $D>1$; it may be that when the dimension of the branches increases, that the number of them no longer scales exponentially. In a recent paper, Stern et al \cite{stern2017} studied examples with quadrilateral faces where $V_e$ was large and $D=0$. They studied branches defined to be minimal-energy directions (as opposed to zero-energy, as in this work) away from the unfolded state. Intriguingly, the scaling of their $B$ is exponential in $\sqrt{V_i}$. It would be very interesting to understand the crossover between the regime studied in that work and this one. 

Proving that each sign-related pair of vertex sign patterns corresponds to a unique pair of folding branches remains elusive. Our problem of finding all branches fits into the broader context of enumeration problems in geometric rigidity such as finding all realizations of isostatic graphs in the plane \cite{Borcea2004,capco2017} and enumerating all rigid clusters of sticky hard spheres \cite{holmes2016enumerating}. The vertex sign patterns seem to impose some structure on the branches in configuration space that leads to the much more tractable formula $B=2^{V_i}$ than e.g.~the recent recursion formula for the number of complex realizations of isostatic graphs in the plane \cite{capco2017}.

Understanding the relation between vertex sign patterns and the geometry of the branches better may be useful for developing robust self-folding structures. In particular it would be interesting to see whether a self-folding paradigm based not only on preferred dihedral angles but also on vertex popping states could be easier to control experimentally, \textit{e.g.}~with conical indenters above and below the sheet, or perhaps with some kind of actuation or swelling that breaks the up-down symmetry of the sheet near each vertex. 

Finally, since MV assignments seem to be more frequently ambiguous, particularly for origami with many internal vertices (see Table \ref{tab:summary} where the fraction of crease patterns with branches having duplicate MV assignments seems to be increasing), we suggest that perhaps origami crease patterns for folding paper origami should be given with vertex popping states as well.

\acknowledgments{
We thank the kind hospitality of the Kavli Instutute of Theoretical Physics in Santa Barbara, CA and thank Tom Hull, Louis Theran, Menachem Stern and Arvind Murugan for helpful discussions. BGC thanks ICERM in Providence, RI and the participants and staff during the 2016 ``Topology in Motion'' semester program there for support and encouragement. This research was supported in part by the National Science Foundation under Grant No.~NSF PHY-1125915, and also Grant No.~EFRI ODISSEI-1240441.}

\appendix

\section{Deriving a quadratic constraint for each internal vertex from the vanishing of the Gaussian curvature}\label{app:gc}

\begin{figure}%[b]
\includegraphics[width=3.5in]{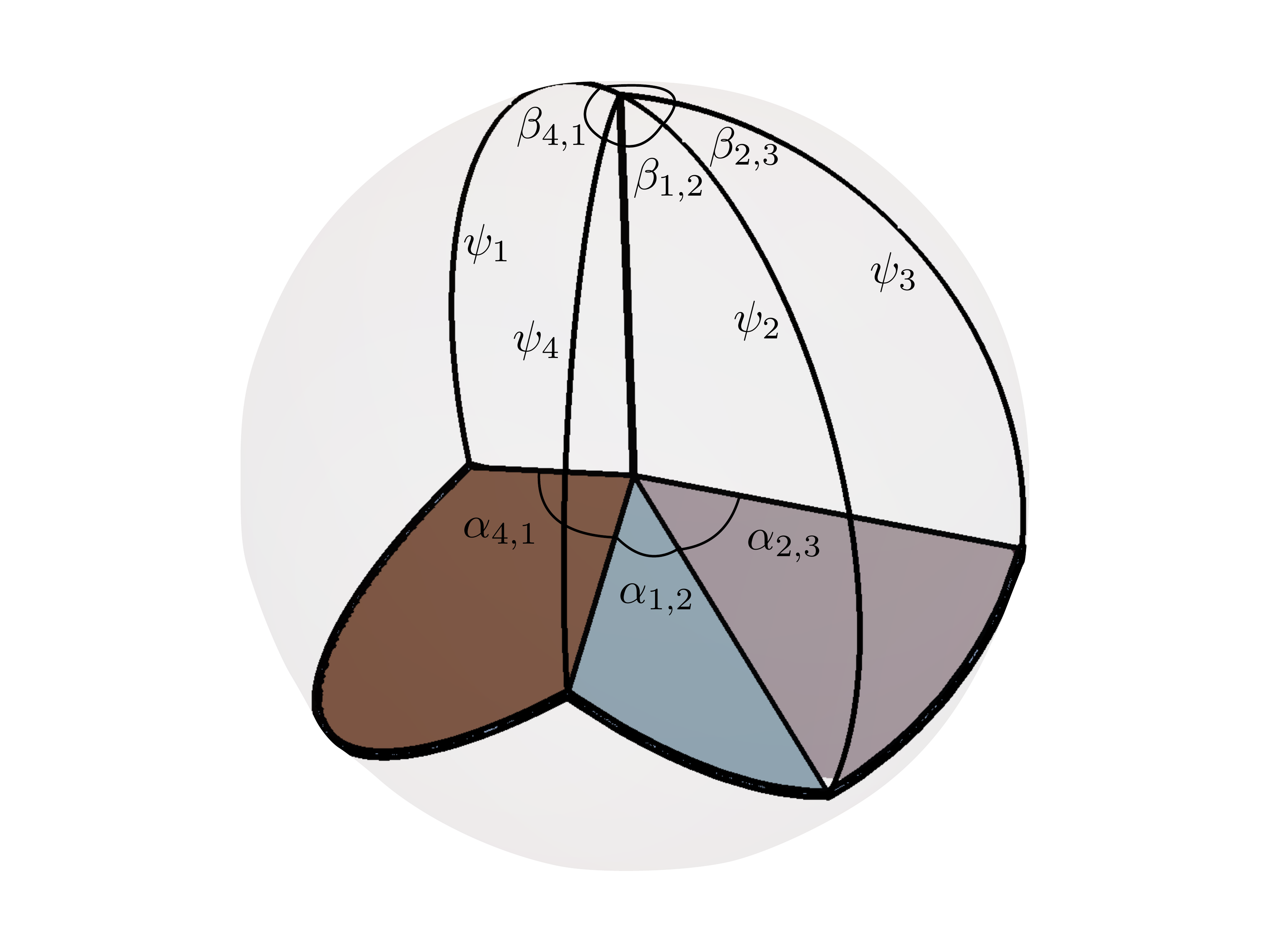}
\caption{\label{fig:singlevertex} The spherical polygon spanned by a single vertex. The angle between a fold and the $\hat{\mathbf{z}}$ axis is $\psi_i$, the planar angle between adjacent folds in the plane is $\alpha_{i,i+1}$ while the angle between adjacent folds at the north pole is $\beta_{i,i+1}$. Note that $\sum_i \beta_{i,i+1} = 2 \pi$.}
\end{figure}

One way to express the Gaussian curvature constraint around a single vertex is in terms of the interior angles of the spherical triangle made by a pair of adjacent folds and the $\hat{\mathbf{z}}$ axis (Fig.\ \ref{fig:singlevertex}). We denote $\alpha_{i,i+1}$ as the planar angle between adjacent folds, which becomes the length of one side of the triangle. Similarly, we denote $\psi_i$ as the angle between $\hat{\mathbf{z}}$ and the $i^{th}$ fold.

Let $h_0$ be the height of the central vertex, and $h_i$ be the heights of the vertices around the interior vertex at $h_0$ and $L_i$ are the lengths of the folds from $h_0$ to $h_i$. We assume that all quantities are periodic in the index and numbered in counterclockwise order. Then we have angles,
\begin{equation}
\psi_i = \frac{\pi}{2}-\frac{h_i-h_0}{L_i}.
\end{equation}

Finally, define the interior angle of the spherical triangle between $\hat{\mathbf{z}}$ and the two folds as $\beta_{i,i+1}$ (Fig.\ \ref{fig:singlevertex}). The relationship between these angles is given by the spherical law of cosines,
\begin{equation}
\cos \alpha_{i,i+1} = \cos \psi_i \cos \psi_{i+1} + \sin \psi_i \sin \psi_{i+1} \cos \beta_{i,i+1}.
\end{equation}

Since the vertices we consider are nearly unfolded, we will expand around an unfolded configuration in which $\psi_i = \pi/2$ and $\beta_{i,i+1} = \alpha_{i,i+1}$. Therefore, if $\psi_i = \pi/2 + \delta \psi_i$, $\beta_{i,i+1} = \alpha_{i,i+1} + \delta \beta_{i,i+1}$, expanding to quadratic order yields
\begin{equation}
\delta \beta_{i,i+1} = \frac{2 \delta \psi_i \delta \psi_{i+1} - \delta \psi_i^2 \cos \alpha_{i,i+1} - \delta \psi_{i+1}^2 \cos \alpha_{i,i+1}}{2 \sin \alpha_{i,i+1}}.
\end{equation}

The sum of the interior angles $\sum_{i} \alpha_{i,i+1} = 2 \pi - K$, where $K$ is the deficit angle, equivalent to the discrete Gaussian curvature of the vertex; however, $\sum_i \beta_{i,i+1} = 2 \pi$ no matter $K$. Therefore, for small $K$, we require that
\begin{equation}
\sum_i \delta \beta_{i,i+1} = K.
\end{equation}
When the origami is unfolded, $K = 0$ at each vertex.

Now we can rewrite the angle $\delta \beta_{i,i+1}$ in terms of the heights as
\begin{eqnarray}
\delta \beta_{i,i+1} &=& -\csc \alpha_{i,i+1} \frac{\Delta L_{i,i+1}}{L_i L_{i+1}} \frac{\left(h_i - h_{i+1}\right)^2}{2 \Delta L_{i,i+1}}\nonumber\\
& & + \left(\frac{\csc \alpha_{i,i+1}}{L_{i+1}} - \frac{\cot \alpha_{i,i+1}}{L_i}\right) \frac{\left(h_i-h_0\right)^2}{2 L_i} \\
& & + \left(\frac{\csc \alpha_{i,i+1}}{L_i} - \frac{\cot \alpha_{i,i+1}}{L_{i+1}}\right) \frac{\left(h_{i+1}-h_0\right)^2}{2 L_{i+1}}, \nonumber
\end{eqnarray}
where $\Delta L_{i,i+1} = \sqrt{L_i^2 + L_{i+1}^2 - 2 L_i L_{i+1} \cos \alpha_{i,i+1}}$ is the distance between the vertices at $h_i$ and $h_{i+1}$.

Finally, we have
\begin{eqnarray}
0 &=& \sum_i -\csc \alpha_{i,i+1} \frac{\Delta L_{i,i+1}}{L_i L_{i+1}} \frac{\left(h_i - h_{i+1}\right)^2}{2 \Delta L_{i,i+1}} \nonumber\\  \label{constraint}
& & + \left(\frac{\csc \alpha_{i,i+1}}{L_{i+1}} + \frac{\csc \alpha_{i-1,i}}{L_{i-1}}\right. \\
& & \left. - \frac{\cot \alpha_{i,i+1}}{L_i} - \frac{\cot \alpha_{i-1,i}}{L_{i}}\right) \frac{\left(h_i-h_0\right)^2}{2 L_i}\nonumber
\end{eqnarray}
This is precisely the form of the quadratic constraint that we get from the self stresses in Eq.~(\ref{eq:compat}).  In the next appendix, we verify that the coefficients here do correspond to a self stress.

\section{Verification that the Gaussian curvature constraint yields a self stress}\label{app:gcwheelstress}

Using Eq.\ (\ref{constraint}), we can immediately read off the components of the stress that would give the quadratic constraint in Eq.~(\ref{constraint}). In particular, the stress in the edges along the ``rim'' of the wheel should be
\begin{equation}\label{eq:outer}
\sigma_{i,i+1} = -\csc \alpha_{i,i+1} \frac{\Delta L_{i,i+1}}{L_i L_{i+1}} 
\end{equation}
and on the spoke edges
\begin{eqnarray}\label{eq:spokes}
\sigma_{i} &=&  \frac{\csc \alpha_{i,i+1}}{L_{i+1}} + \frac{\csc \alpha_{i-1,i}}{L_{i-1}} \\
& & - \frac{\cot \alpha_{i,i+1}}{L_i} - \frac{\cot \alpha_{i-1,i}}{L_{i}}.\nonumber
\end{eqnarray}

What remains to to verify that these do in fact satisfy the equation $\mathbf{\sigma}^T\cdot\mathbf{C}=\mathbf{0}^T$ defining self stresses. Recall that self stresses are assignments of tensions and compressions to edges that preserve force balance at each vertex. Therefore we must check force balance at the spoke vertices (labeled $i=1$ to $N$) and the hub vertex 0.  

\subsection{Force balance at the spokes}

\begin{figure}%[b]
\includegraphics[width=3.5in]{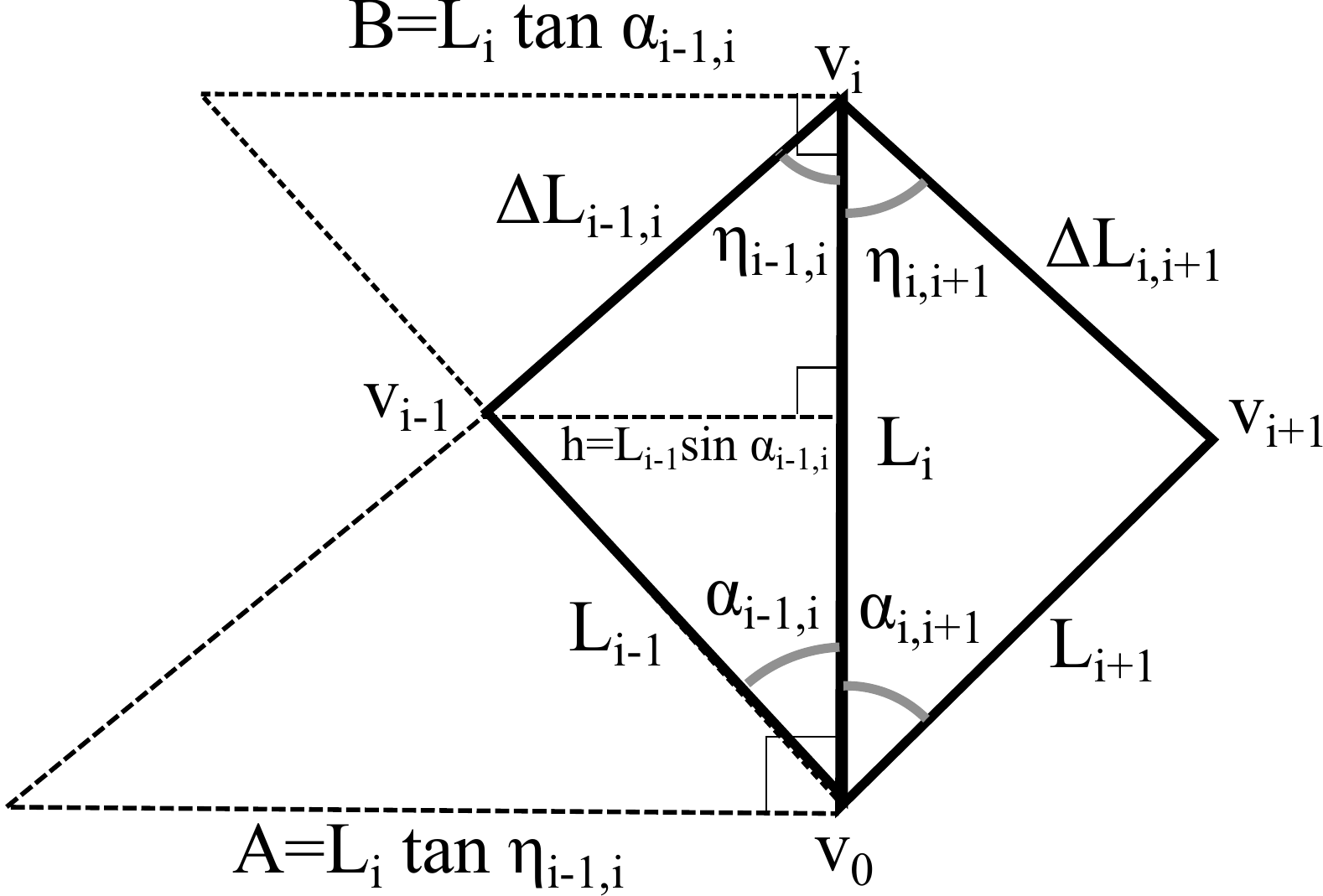}
\caption{\label{fig:crossedladder} Trigonometric diagram for verification of force balance at spoke vertex $i$ along the direction parallel to the spoke. The ``crossed ladder'' identity relates the lengths of the three parallel dashed lines, $1/A+1/B=1/h$.}
\end{figure}

We first show force balance at spoke vertex $i$.  Recall that the position of vertex $j$ is the vector $\mathbf{U}_j$ so that $L_i=|\mathbf{U}_i-\mathbf{U}_0|$ and $\Delta L_{i,i+1}=|\mathbf{U}_{i+1}-\mathbf{U}_i|$ (Fig.~\ref{fig:crossedladder}). Let's first check the forces in the direction perpendicular to the spoke edge vector ($\mathbf{U}_i - \mathbf{U}_0$).  We only need to use the stresses along the outer edges (Eq.\ (\ref{eq:outer})) for this.

First we use the law of sines on the triangle with sides $L_i, L_{i+1}, \Delta L_{i,i+1}$ to transform Eq.\ (\ref{eq:outer}) to 
\begin{equation}\label{eq:outer2}
\sigma_{i,i+1} = - \csc \eta_{i,i+1} / L_{i}, 
\end{equation}
where $\eta$ is the angle opposite $L_{i+1}$.  A similar argument shows that 
\begin{equation}\label{eq:outer3}
\sigma_{i-1,i} = -\csc \eta_{i-1,i} / L_i,
\end{equation}
where $\eta_{i-1,i}$ is opposite $L_{i-1}$.

The magnitudes of the forces perpendicular to the spoke vector are given by $\sigma_{i,i+1}\sin\eta_{i,i+1}=-\sigma_{i-1,i}\sin\eta{i-1,i}$ so there is force balance along this direction at vertex $i$.

Next we consider the force at vertex $i$ in the direction parallel to the spoke vector.  Using Eqs.\ (\ref{eq:outer2}) and (\ref{eq:outer3}), we see that the contribution from the two rim edges to the force parallel to the spoke vectors may be written
\begin{align}
F_{rim,\parallel}&=-(\cot\eta_{i-1,i} + \cot\eta_{i,i+1}) / L_i.\\
&=F_{rim,\parallel,-}+F_{rim,\parallel,+}\nonumber\\
F_{rim,\parallel,-}&=-\cot\eta_{i-1,i}/L_i\label{eq:spokepar}\\
F_{rim,\parallel,+}&=-\cot\eta_{i,i+1}/L_i
\end{align}

Consider the term $F_{rim,\parallel,-}$. This is minus the reciprocal of $L_i \tan\eta_{i-1,i}$ which is the side length $A$ of a certain right triangle with $L_i$ as one of its legs (Fig.~\ref{fig:crossedladder}). There is a similar interpretation for the term $F_{rim,\parallel,+}$.

Now consider the contribution from the spoke edge given by Eq.\ (\ref{eq:spokes}). We will split this expression into two pieces, which will each cancel one of the two terms $F_{rim,\parallel,\mp}$. The first piece is
\begin{equation}\label{eq:firstpiece}
\csc\alpha_{i-1,i} / L_{i-1} - \cot\alpha_{i-1,i} / L_i .
\end{equation}

The first term of Eq.~(\ref{eq:firstpiece}) is the reciprocal of $L_{i-1}\sin\alpha_{i-1,i}$, which is the side length $h$ of a right triangle with $L_{i-1}$ as the hypotenuse. The second term is the reciprocal of $L_i \tan\alpha_{i-1,i}$ which is the side length $B$ of a (different) right triangle with $L_i$ as one of its legs. See Fig.~\ref{fig:crossedladder} where these sides are depicted as dashed lines.

The three side lengths $A,B,h$ are related as in the ``Crossed Ladders problem'' \cite{crossedladders}: for parallel line segments $A,B,h$ in the configuration shown in Fig.~\ref{fig:crossedladder}, we have the identity $1/A + 1/B = 1/h$. Therefore the first part of the force along the spoke edge (Eq.~(\ref{eq:firstpiece})) cancels $F_{rim,\parallel,-}$ (Eq.~(\ref{eq:spokepar})). 

It should be clear that the argument of the preceding three paragraphs can also be carried out for the second piece of Eq.\ (\ref{eq:spokes}), to show that it cancels $F_{rim,\parallel,+}$.

\subsection{Force balance at the hub}

For force balance at the hub vertex, we proceed with an induction on the number of spokes. For the base case with 3 spokes, we must show that the vectors with lengths $\sigma_1,\sigma_2,\sigma_3$ directed along the spoke directions sum to zero. Let $\mathbf{u}_j=(\mathbf{U}_j-\mathbf{U}_0)/|\mathbf{U}_j-\mathbf{U}_0|$ be the unit vector along the $j$th spoke edge. Then this is:
\begin{equation}
\sum_{j=1}^3\sigma_j\mathbf{u}_j=\mathbf{0}.
\end{equation}

We interpret this equation as the condition for the closure of the polygonal path with sides $\sigma_j\mathbf{u}_j$. Generally, when a vertex is in equilibrium, the forces acting on it can be seen as the sides of a closed {\em force polygon}. Note that in this force triangle the angles $\alpha$ between the spoke edges become the ``turning angles'', so that e.g.~the angle opposite side $\sigma_1$ in the force triangle is $\pi-\alpha_{2,3}$. We can show that this triangle is closed using the (converse of the) law of sines, which amounts to proving that the following are all equal:
\begin{equation}
\sigma_1/\sin\alpha_{2,3}=\sigma_2/\sin\alpha_{3,1}=\sigma_3/\sin\alpha_{1,2}
\end{equation}

We will just show the first equality, as the proof of the others is exactly the same.
\begin{align}
\frac{\sigma_1}{\sin\alpha_{2,3}}=&\csc\alpha_{2,3}\left(\frac{\csc \alpha_{1,2}}{L_{2}} + 
\frac{\csc \alpha_{3,1}}{L_{3}}\right. \\
& \left. - \frac{\cot \alpha_{1,2}}{L_1} - \frac{\cot \alpha_{3,1}}{L_{1}}\right)\nonumber\\
=&\frac{\csc\alpha_{3,1}\csc\alpha_{2,3}}{L_3}+\frac{\csc\alpha_{3,1}\csc\alpha_{1,2}}{L_1}\\
&+\frac{-\csc\alpha_{3,1}(\cot\alpha_{2,3}+\cot\alpha_{1,2})}{L_2}.\nonumber
\end{align}

We will need the following trigonometric identity, equivalent to the sine addition rule:
\begin{align}
%\sin(a+b)=&\sin a\cos b+\cos a\sin b\\
%\frac{\sin(a+b)}{\sin a\sin b}=&\cot b+\cot a\\
\csc a\csc b=&\csc(a+b)(\cot a+\cot b).\label{eq:simpletrig}
\end{align}
Note that $-\csc\alpha_{3,1}=\csc(\alpha_{1,2}+\alpha_{2,3})$ since $\alpha_{1,2}+\alpha_{2,3}+\alpha_{3,1}=2\pi$. Therefore by applying Eq.\ (\ref{eq:simpletrig}) twice, we have,
\begin{align}
\frac{\sigma_1}{\sin\alpha_{2,3}}=&\frac{\csc\alpha_{3,1}\csc\alpha_{2,3}}{L_3}+\frac{-\csc\alpha_{2,3}(\cot\alpha_{1,2}+\cot\alpha_{2,3})}{L_1}\\
&+\frac{\csc\alpha_{2,3}\csc\alpha_{1,2}}{L_2}\nonumber\\
%=&\csc\alpha_{3,1}\left(\frac{\csc \alpha_{2,3}}{L_{3}} + \frac{\csc \alpha_{1,2}}{L_{1}}\right. \\
%& \left. - \frac{\cot \alpha_{2,3}}{L_2} - \frac{\cot \alpha_{1,2}}{L_{2}}\right)\nonumber\\
=&\frac{\sigma_2}{\sin\alpha_{3,1}}.
\end{align}

\begin{figure}%[b]
\includegraphics[width=3.5in]{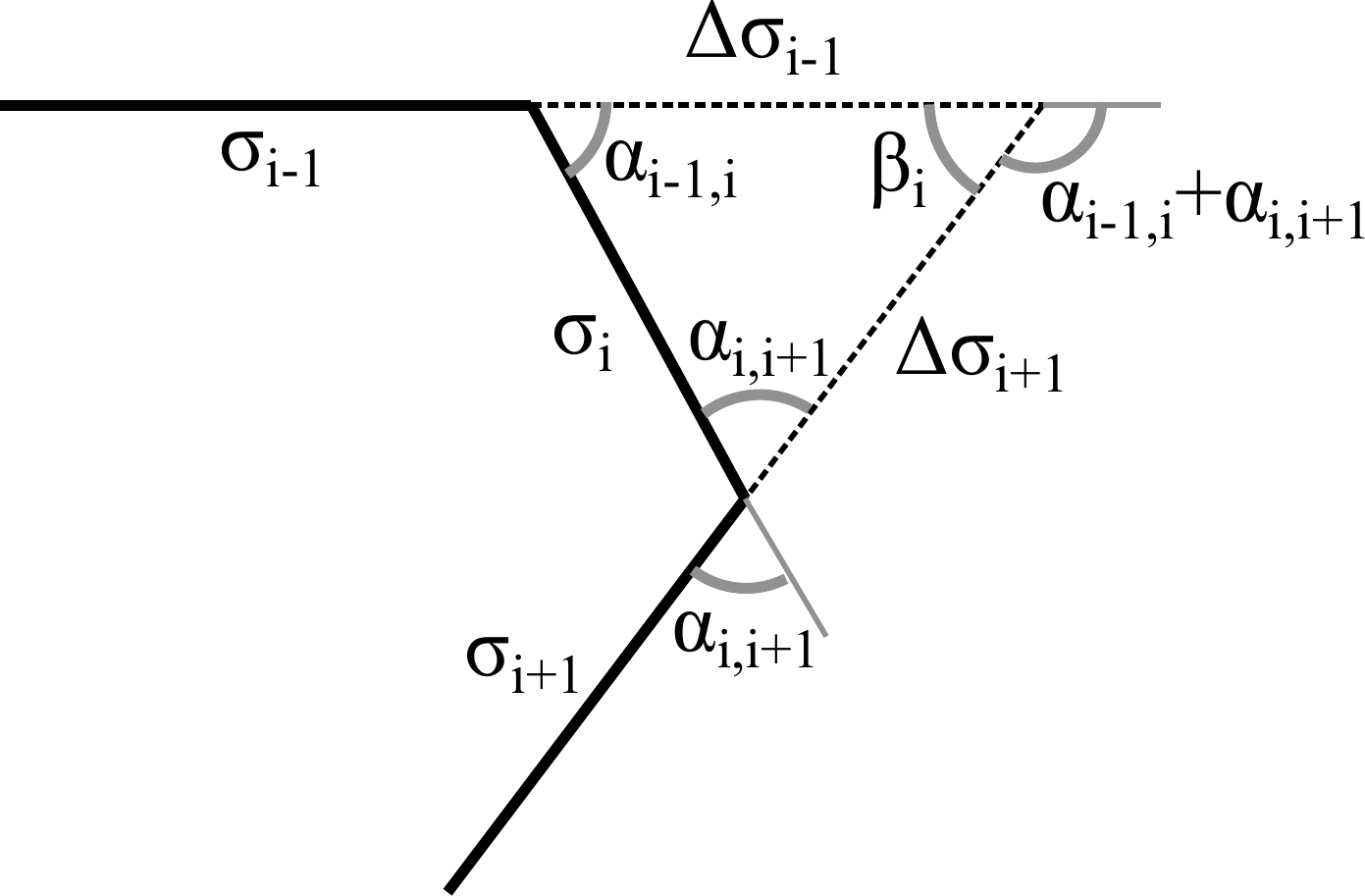}
\caption{\label{fig:hubbalance} Portion of the force polygon for the hub vertex considered during the induction. Force balance at the hub vertex is equivalent to the closure of the force polygon. Each edge in the figure is a vector parallel to the spoke edges whose length is equal to the magnitude of the stress component in that edge. Here we only show the forces along spokes $i-1,i,i+1$ and how they change if spoke $i$ were to be removed.}
\end{figure}

For the induction step, assume that we have shown that we have force balance at the hub vertex in any wheel graph with $n$ spokes with stresses given by Eq.~(\ref{eq:spokes}).  Consider the star subgraph $G_{n+1}$ of the wheel graph, consisting of $n+1$ spoke vertices connected to the hub; we will reduce the force balance at its hub to force balance in the star graph $G_{n,i}$ formed by removing (an arbitrary) spoke vertex $i$ and the edge joining it to the hub.  As in the argument for the base case, we will work with force polygons and prove that the closure of the force polygon $P_{n,i}$ of $G_{n,i}$ implies the closure of the force polygon $P_{n+1}$ of $G_{n+1}$. 

Note that the expression in Eq.~(\ref{eq:spokes}) for spoke $j$ depends only on the lengths $L_{j-1},L_j,L_{j+1}$ and the angles $\alpha_{j-1,j}$ and $\alpha_{j,j+1}$.  This means that the self stress evaluated on corresponding spokes of $G_{n+1}$ and $G_{n,i}$ are identical except at the edges $i-1,i,i+1$. Thus $P_{n,i}$ and $P_{n+1}$ are identical except at those edges too. We will prove that the situation is as depicted in Fig.~\ref{fig:hubbalance}. There, the edges of $P_{n+1}$ are depicted as thick lines, with the dashed edges $i-1,i+1$ of $P_{n,i}$ overlaid.  

To get started, we observe that the closure of $P_{n,i}$ implies there is a vertex where the edges corresponding to spokes $i-1$ and $i+1$ (dashed in Fig.~\ref{fig:hubbalance}) intersect. We will denote lengths of edge $i-1$ and $i+1$ in $P_{n,i}$ are $\sigma^0_{i-1}$ and $\sigma^0_{i+1}$, respectively.  In $P_{n+1}$, edges $i-1$ and $i+1$ can be taken to lie on the corresponding edges of $P_{n,i}$ but they will in general have different lengths, which are simply $\sigma_{i-1}$ and $\sigma_{i+1}$. For $P_{n+1}$ to be closed, edge $i$ must begin at the end point of $i-1$ and end at the starting point of $i+1$.  This means that there is a closed triangle with side lengths $\Delta\sigma_{i-1}=\sigma_{i-1}-\sigma^0_{i-1}$, $\Delta\sigma_{i+1}=\sigma_{i+1}-\sigma^0_{i+1}$, and $\sigma_i$ and whose angles are determined by the angles $\alpha_{i-1,i}$ and $\alpha_{i,i+1}$ between the spoke edges $i-1,i$ and $i+1$.  Note that the angle $\beta_i=\pi-\alpha_{i-1,i}-\alpha_{i,i+1}$.

To prove that this triangle is closed, we will again use the law of sines.  For convenience, here are the formulas for the lengths:
\begin{align}
\sigma_i=&\frac{\csc \alpha_{i,i+1}}{L_{i+1}} + \frac{\csc \alpha_{i-1,i}}{L_{i-1}} \\
& - \frac{\cot \alpha_{i,i+1}+\cot\alpha_{i-1,i}}{L_i}\nonumber\\
\Delta\sigma_{i-1}=&\frac{\csc \alpha_{i-1,i}}{L_{i}} + \frac{\csc \alpha_{i-2,i-1}}{L_{i-2}} \\
& - \frac{\cot \alpha_{i-1,i}+\cot\alpha_{i-2,i-1}}{L_{i-1}}\nonumber \\
& -\frac{\csc( \alpha_{i-1,i}+\alpha_{i,i+1})}{L_{i+1}} - \frac{\csc \alpha_{i-2,i-1}}{L_{i-2}} \nonumber\\
& + \frac{\cot (\alpha_{i-1,i}+\alpha_{i,i+1})+\cot\alpha_{i-2,i-1}}{L_{i-1}}\nonumber\\
=&\frac{\csc\alpha_{i-1,i}}{L_i}-\frac{\csc(\alpha_{i-1,i}+\alpha_{i,i+1})}{L_{i+1}}\\
&-\frac{\cot\alpha_{i-1,i}-\cot(\alpha_{i-1,i}+\alpha_{i,i+1})}{L_{i-1}}\nonumber\\
\Delta\sigma_{i+1}=&\frac{\csc\alpha_{i,i+1}}{L_i}-\frac{\csc(\alpha_{i-1,i}+\alpha_{i,i+1})}{L_{i-1}}\\
&-\frac{\cot\alpha_{i,i+1}-\cot(\alpha_{i-1,i}+\alpha_{i,i+1})}{L_{i+1}}.\nonumber
 \end{align}

We will only show $\sigma_i/\sin\beta_i=\Delta\sigma_{i-1}/\sin\alpha_{i,i+1}$, as the proof of the other identity is the same.
\begin{align}
\frac{\sigma_i}{\sin\beta_i}=&-\csc(\alpha_{i-1,i}+\alpha_{i,i+1})\left(\frac{\csc \alpha_{i,i+1}}{L_{i+1}} + \frac{\csc \alpha_{i-1,i}}{L_{i-1}}\right. \\
&\left. - \frac{\cot \alpha_{i,i+1}+\cot\alpha_{i-1,i}}{L_i}\right)\nonumber\\
=&-\csc(\alpha_{i-1,i}+\alpha_{i,i+1})\left(\frac{\csc \alpha_{i,i+1}}{L_{i+1}} + \frac{\csc \alpha_{i-1,i}}{L_{i-1}}\right) \\
& + \frac{\csc\alpha_{i,i+1}\csc\alpha_{i-1,i}}{L_i}\nonumber\\
=&\csc\alpha_{i,i+1}\left(\frac{-\csc(\alpha_{i-1,i}+\alpha_{i,i+1})}{L_{i+1}} 
 + \frac{\csc\alpha_{i-1,i}}{L_i}\right)\\
&- \frac{\csc \alpha_{i,i+1}(\cot\alpha_{i-1,i}-\cot(\alpha_{i-1,i}+\alpha_{i,i+1}))}{L_{i-1}} \nonumber\\
=&\frac{\Delta\sigma_{i-1}}{\sin\alpha_{i,i+1}}.
\end{align}
where in the second and third equalities we have applied Eq.~(\ref{eq:simpletrig}).  This completes the proof that Eqs.~(\ref{eq:outer}) and (\ref{eq:spokes}) define a self stress on the wheel graph.

\section{Vertex Quadratic Forms}\label{app:quadratic}
Let us consider the quadratic form in Eq.~(\ref{constraint}). This form has the interpretation as the energy if the wheel self stress is applied as a pre-stress \cite{connellywhiteley}.  Note that with our sign convention, this pre-stress places the spokes under compression and the outer edges under tension.

Using the same notation as in Section \ref{sec:methods}, where $h_0$ is the height of the vertex and $h_1$ through $h_N$ are the heights of the adjacent vertices, Eq.~(\ref{constraint}) then reads
\begin{eqnarray}
0 &=& \sum_{n=1}^N \left(g_{n,n+1} 2 h_n h_{n+1} + f_n h_n^2 + A_n 2 h_0 h_n \right) - h_0^2 \sum_{n=1}^N A_n\nonumber,
\end{eqnarray}
where 
\begin{align}
f_n =& \frac{\cot \alpha_{n,n+1} + \cot \alpha_{n-1,n}}{L_n^2}\\
g_{n,n+1} =& -\frac{\csc \alpha_{n,n+1}}{L_n L_{n+1}}\\
A_n=&-f_n-g_{n,n+1}-g_{n-1,n}=\frac{\sigma_n}{L_n}.
\end{align}
The corresponding matrix has three null directions, corresponding to the vertical translation and rotations about the $x-$ and $y-$axes of the entire origami structure. We can remove these by setting $h_0 = h_1 = h_N = 0$ explicitly. Then we have
\begin{equation}\label{form}
0 = \sum_{n=2}^{N-1} \left(g_{n,n+1} 2 h_n h_{n+1} + f_n h_n^2 \right).
\end{equation}
This gives a tridiagonal matrix whose determinant, $\chi(N)$, can be computed to be
\begin{equation}\label{determinant}
\chi(N) = \frac{\prod_{n=1}^{N-1} \csc \alpha_{n,n+1}}{\prod_{n=2}^{N-1} L_n^2} \sin \left(\sum_{n=1}^{N-1} \alpha_{n,n+1}\right).
\end{equation}
This can be proven by induction using the continuant,
\begin{equation}
\chi(N+2) = f_{N+2}\chi(N+1) - g_{N+1,N+2}^2 \chi(N).
\end{equation}
% check these indices again?

When all the angles between vertices are identical, $\alpha_{n,n+1} \equiv \alpha$ and the lengths of the folds are also the same, $L_n = L$, the quadratic form of Eq.~(\ref{form}) is also Toeplitz, and its eigenvalues can be determined explicitly by
\begin{equation}
\lambda_m = f + 2 |g| \cos \left( \frac{m \pi}{N-1} \right)
\end{equation}
for $m$ ranging from $1$ to $N-1$. For $N$ folds, $\alpha = 2 \pi/N$. Thus,
\begin{equation}\label{eq:equalangleeigs}
\lambda_m = 2 \frac{\csc \left(2 \pi/N \right)}{L^2} \left[ \cos \left( \frac{2 \pi}{N} \right) -  \cos \left( \frac{m \pi}{N-1} \right) \right].
\end{equation}
Consequently, $\lambda_1 < 0$ and the other eigenvalues are positive.  Now consider changing the angles smoothly.  Since no eigenvalue can change sign without the determinant $\chi = 0$, we see that Eq.~(\ref{determinant}) implies that there is always one negative eigenvalue so long as the angles between any pair of adjacent folds remain between 0 and $\pi$. When $N\geq4$ there are thus eigenvalues of both signs, and there is always a solution to Eq.~(\ref{form}) for a single vertex; the vertex is infinitesimally rigid when all eigenvalues have the same sign (as happens when $N=3$, for instance).

We can get some physical intuition for the distribution of eigenvalues as follows. Under the wheel pre-stress corresponding to this quadratic energy, the eigenvector with negative eigenvalue should correspond to an out-of-plane displacement that is maximally unstable. We can imagine that there is one that increases the lengths of the compressed spokes while not increasing the lengths of the outer edges too much, as such a motion will be destabilized by the stress.  The other $N-3$ motions correspond to out-of-plane displacements that are stabilized by the pre-stress, i.e.~those that primarily increase the lengths of the outer edges.

The discussion above recovers a special case of a result of Kapovich and Millson \cite{kapovichmillson} who studied the configuration space of origami vertices using techniques from the deformation theory of representations of $SO(3)$.  In fact, they considered unfolded origami vertices that may fold back on themselves (allowing $\alpha_{j,j+1}<0$ for some $j$) or have a different winding around the vertex ($\sum_{j=1}^N\alpha_{j,j+1}=2\pi w$, for some integer $w$ not necessarily equal to 1).  The relevant result of their paper is the following

\begin{thm}[Kapovich and Millson, 1997]\label{thm:km}
Theorem 1.1(iii). Assume all planar angles satisfy $0<|\alpha_{j,j+1}|<\pi$. The germ of the configuration space of an origami vertex with $f$ forward-tracks, $b$ back-tracks, and winding $w$ is isomorphic to the germ of the null-cone of a quadratic form with nullity 3, and signature $(f-2w-1,b+2w-1)$.
\end{thm}

Here $f$ counts the number of forward-tracks, defined to be the planar angles with $\alpha>0$ and $b$ counts the number of back-tracks, defined to be those angles with $\alpha<0$. 

We showed above the case of this theorem when $f=N$, $b=0$ and $w=1$, and in fact our sign convention for the sign of the quadratic form also agrees with theirs.  (A change of sign swaps negative and positive eigenvalues, but of course leads to the same null-cone).  Furthermore, it is easy to see that our observation that Eq.~(\ref{determinant}) only vanishes when one of the $\alpha$ is an integer multiple of $\pi$ is consistent with their statement that the signature changes when any of $f,b,w$ change.  Indeed, Eq.~(\ref{form}) for the quadratic form applies in full generality, and so we can actually use it to give a complete proof of Theorem \ref{thm:km}.

{\bf Proof of Theorem \ref{thm:km}}: The first claim about the germ being isomorphic to the germ of the null-cone essentially states that the quadratic form we have derived in Eq.~(\ref{form}) as the lowest order constraint on the configuration space does give an accurate picture of a neighborhood of the singular point (the unfolded state), i.e.~that the second-order motions satisfying Eq.~(\ref{form}) extend to true motions.  For this we refer to Theorem 4 of \cite{reciprocalarea} which gives an elementary proof of this fact. (In fact, \cite{kapovichmillson} prove the stronger result that there is an analytic isomorphism between neighborhood germs.)

The rest of the theorem addresses the signature of the quadratic form. The nullity of 3 corresponds to the global isometries mentioned earlier.  So we just need to derive the expression $(f-2w-1,b+2w-1)$ for the number of (positive, negative) eigenvalues.  Since the defining symmetric matrix is $(n+1)\times(n+1)$ and $f+b=n$ it is enough to show that the number of negative eigenvalues $N_-=b+2w-1$.  

Suppose we have a real symmetric $k\times k$ matrix $M$, and let $\Delta_j$ is the determinant of the upper-left $j\times j$ submatrix of $M$ (the $j$th principal minor of $M$).  The key observation is that, provided none of the $\Delta_j$ vanish, $N_-$ is equal to the number of sign changes of the sequence $\Delta_0=1,\Delta_1,\dots,\Delta_k$ \cite{signaturereference}. 

Due to the tridiagonal nature of the stress matrix, $\chi(j)$ is also an expression for $\Delta_j$ of the stress matrix.  So we just have to show that we get exactly $b+2w-1$ sign changes. Let us consider the ratio:
\begin{equation}\label{ratio}
\frac{\Delta_{j+1}}{\Delta_j} = \frac{\csc \alpha_{j-1,j}}{L_j^2}\frac{ \sin \left(\sum_{n=1}^{j} \alpha_{n,n+1}\right)}{ \sin \left(\sum_{n=1}^{j-1} \alpha_{n,n+1}\right)}.
\end{equation}
This consists of two factors.  The first factor $\csc\alpha_{j-1,j}$ is negative if and only if $\alpha_{j-1,j}<0$, i.e.~when the $j$th planar angle is a back-track.  The second factor is negative if and only if the two quantities $\sum_{n=1}^{j-1}\alpha_{n,n+1}$ and $\sum_{n=1}^{j}\alpha_{n,n+1}$ sandwich an integer multiple of $\pi$.  Let us first assume that the second factor does not vanish (the first never will, by our assumptions on the $\alpha$'s).

We get a net negative sign in Eq.~(\ref{ratio}) if and only if one of the following two possibilities occurs. Possibility A, the $j$th planar angle is a back-track and the partial sum $\sum_{n=1}^j\alpha_{n,n+1}$ does not pass an integer multiple of $\pi$. If $C_b$ is the number of back-track crossings over integer multiples of $\pi$, this occurs $b-C_b$ times. Possibility B, the $j$th planar angle is a forward-track and the partial sum does cross an integer multiple of $\pi$. This occurs $C_f$ times, where $C_f$ is the number of forward-track crossings of integer multiples of $\pi$.

It follows that $2w-1$ is the {\em net} number of times $C_f-C_b$ that integer multiples of $\pi$ are crossed, since $w$ is the total winding number. Therefore we have $b-C_b+C_f=b+2w-1$ sign changes in total, as desired.

Finally, we treat the case where $\Delta_j$ vanishes due to the partial sum $\sum_{n=1}^j\alpha_{n,n+1}$ being equal to an integer multiple of $\pi$.  However, as argued after Eq.\ (\ref{eq:equalangleeigs}), we can perturb the angles $\alpha$ to avoid these cases without changing the sign of the overall determinant (note that the angle sum in $\chi(N)$ is equal to $2\pi-\alpha_{N,1}$), and hence without changing the signature. Such a perturbation can also be chosen small enough so that $f,b,w$ do not change, so the same formula applies.

This concludes the proof of Theorem \ref{thm:km}.

\section{High-dimensional geometry and random origami}\label{app:highD}

\begin{figure}
\includegraphics[width=3.5in]{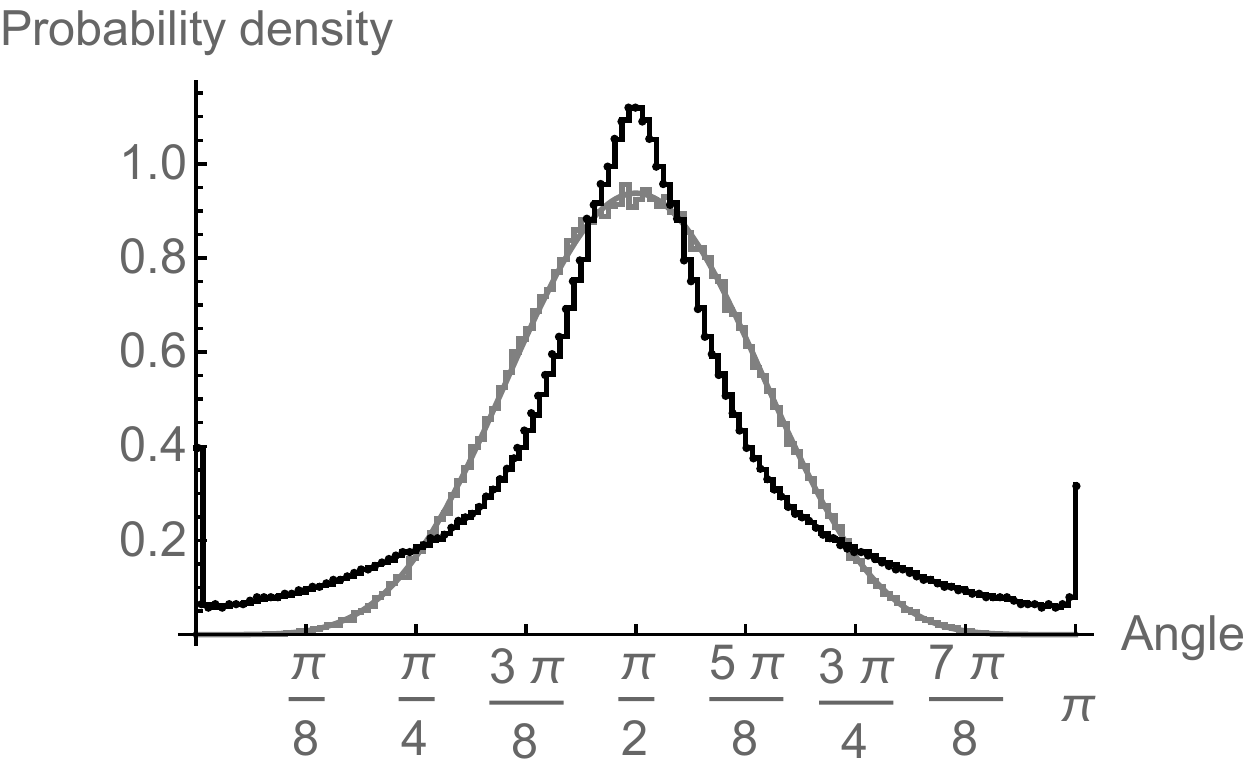}
\caption{\label{fig:v6} The probability distribution for the angles between branch vectors in random origami with $V_i = 6$ (black) compared to randomly distributed points on a 6-dimensional sphere (gray). The bin sizes are $\pi/32$. The solid gray line is an analytical prediction using Eq.\ 5 of Ref.\ \cite{caifanjiang}.}
\end{figure}

To better understand how the branches of a random triangulated $D=1$ origami are distributed as directions in configuration space, we computed pair distribution functions between the branches, where the branches are intepreted as lines in the $3V_i+1$-dimensional space of infinitesimal changes in dihedral angle. (Note that the branches lie in a $V_i+1$-dimensional linear subspace, as not all sets of dihedral angles are induced by height displacements). Computationally, given a particular configuration, we take the numerically computed branches given as vectors of vertex heights and transform them into vectors of dihedral angles. Here we choose to work with both $\pm$ ends of the branch line, so that we have $2^{V_i+1}$ vectors. We then compute the $(2^{V_i+1})^2$ angles coming from dot products between all ordered pairs of branches. These angles are the distances between branches as points on the $V_i$-dimensional sphere.  For each $V_i$ we compute these angles for all of the configurations in Table \ref{tab:summary} to create histograms of the angles, or pair distribution functions.

In Fig.\ \ref{fig:v6}, we show the histogram of the angle between pairs of branches for $V=6$ using bins of size $\pi/128$ (black line).  We compare the results to random points on a $V_i$-dimensional sphere (gray line). The data shows that random origami has a slight enhancement in the number of orthogonal branches but a more prominent enhancement at the two tails of the distribution.  In particular, the results of Ref.~\cite{caifanjiang} imply that for random points on a sphere, the angle distribution approaches a Gaussian with variance going to zero. However, the tails of the distributions of angles between branches in random origami appear to decay exponentially, rather than as a Gaussian. 

\begin{figure}
\includegraphics[width=3.5in]{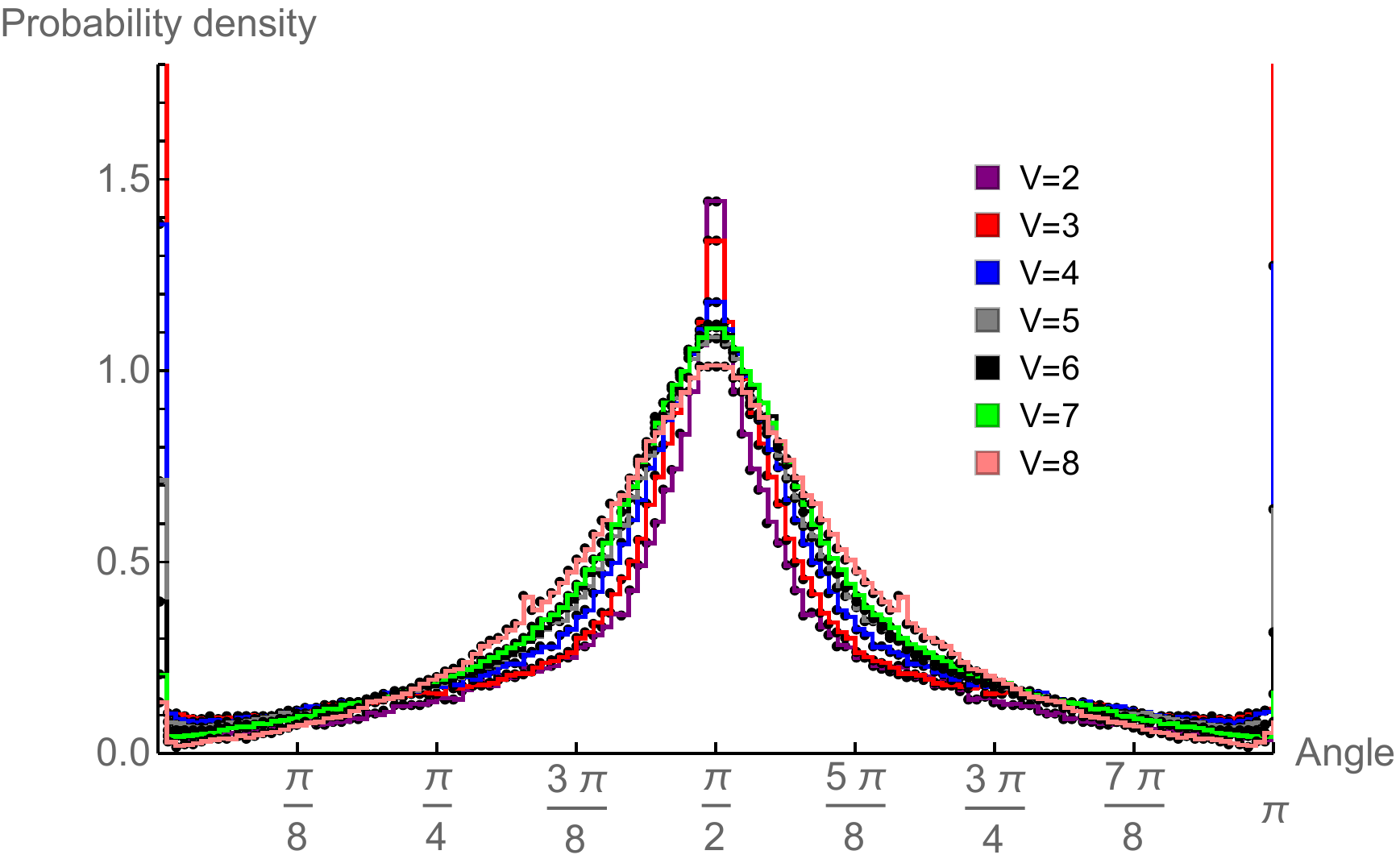}
\caption{\label{fig:distribution} (color) The probability distribution that any two pairs of branches will be a given angle apart for random origami with $V=2$ -- $8$ vertices. The bin sizes are $\pi/32$.}
\end{figure}

To indicate what happens as $V_i$ changes, Fig.\ \ref{fig:distribution} depicts the distributions for random origami with vertices from $V_i=2$ -- $8$, again binning the results into bins of size $\pi/128$. Since the branches have no natural orientation, for each angle $\theta$, we also include the angle $\pi - \theta$; as a consequence, for small $V_i$, there is an enhancement for branches that are almost $\pi$ apart in angle, since each branch has an angle $0$ and $\pi$ with itself. It is interesting to note that while the distributions for $V_i=6,7$ very nearly coincide, they differ significantly from that of $V_i=8$. We do not believe that this is due to a lack of data. Even though we computed fewer configurations at $V_i=8$, each configuration has twice the number of branches and hence four times as many angle pairs, so the number of data points going into the histograms is comparable.

\bibliography{RandomOrigami}

\end{document}